\documentclass{article}

\usepackage{amssymb,amsfonts,amsmath}
\usepackage{cite,enumerate,float}
\usepackage{color}
\usepackage{tikz}
\usetikzlibrary{arrows,snakes,backgrounds}
\usepackage{pdfpages}
\usepackage{hyperref}
\usepackage{authblk}

\def\be{\begin{eqnarray}}
\def\ee{\end{eqnarray}}
\def\nn{\nonumber}

\def\tr{{\rm tr}\,}
\def\Tr{{\rm Tr}\,}

\definecolor{red}{rgb}{1,0,0}
\definecolor{orange}{rgb}{1,0.5,0}
\definecolor{violet}{rgb}{0.7,0,1}

\usepackage{textcomp}
\usepackage{color}
\usepackage{tcolorbox,pict2e}
\usepackage{authblk}
\usepackage{tikz}\usetikzlibrary{tikzmark}\usetikzlibrary{calc}
\usetikzlibrary{arrows,snakes,backgrounds}
\usetikzlibrary{calc, knots, hobby}
\usetikzlibrary{calc, knots, hobby, arrows.meta}
\pgfdeclarelayer{background layer}
\pgfdeclarelayer{foreground layer}
\pgfsetlayers{background layer,main,foreground layer}
\tikzstyle{arrow} = [thick,-{Stealth[length=5mm]}]
\date{}

\hoffset= - 1.05in         

\begin{document}

\title{ \hspace{-2mm} Vogel's universality  and the classification problem for Jacobi identities}

\author{{\bf A. Morozov}\thanks{\href{mailto:morozov@itep.ru}{morozov@itep.ru}} }
\author{{\bf A. Sleptsov}\thanks{\href{mailto:sleptsov@itep.ru}{sleptsov@itep.ru}}}


\affil[]{Moscow Institute of Physics and Technology, 141700, Dolgoprudny, Russia\\
Institute for Information Transmission Problems, 127051, Moscow, Russia\\
NRC "Kurchatov Institute", 123182, Moscow, Russia\footnote{former Institute for Theoretical and Experimental Physics, 117218, Moscow, Russia}}
\renewcommand\Affilfont{\itshape\small}

\maketitle

\vspace{-6.5cm}

\begin{center}
\hfill MIPT/TH-10/25\\
\hfill IITP/TH-12/25\\
\hfill ITEP/TH-10/25
\end{center}

\vspace{3.5cm}

\begin{abstract}
This paper is a summary of discussions at the recent ITEP-JINR-YerPhI workshop on Vogel theory in Dubna. We consider relation between Vogel divisor(s) and the old Dynkin classification of simple Lie algebras. We consider application to knot theory and the  hidden role of Jacobi identities in the definition/invariance of Kontsevich integral, which is the knot polynomial with the values in diagrams, capable of revealing all Vassiliev invariants -- including the ones, not visible  in  other approaches. Finally we comment on the possible breakdown of Vogel universality after the Jack/Macdonald deformation. Generalizations to affine, Yangian and DIM algebras are also mentioned. Especially interesting could be the search for universality in ordinary Yang-Mills theory and its interference with confinement phenomena.
\end{abstract}

\bigskip

\section{Introduction}

\begin{figure}[h] \label{STUfig}
\begin{picture}(100,60)(-140,-20)
\put(0,0){\vector(0,1){20}}
\put(-15,-15){\vector(1,1){15}}
\put(15,-15){\vector(-1,1){15}}
\put(-15,35){\vector(1,-1){15}}
\put(0,20){\vector(1,1){15}}
\put(-23,-15){$a$}
\put(19,-15){$b$}
\put(-23,35){$c$}
\put(19,35){$d$}
\put(-8,7){$e$}
\put(29,5){$-$}
\put(50,25){\vector(1,-1){15}}
\put(50,-5){\vector(1,1){15}}
\put(65,10){\vector(1,-0){20}}
\put(100,-5){\vector(-1,1){15}}
\put(85,10){\vector(1,1){15}}
\put(43,-10){$a$}
\put(102,-10){$b$}
\put(43,25){$c$}
\put(102,25){$d$}
\put(72,13){$e$}
\put(116,5){$+$}
\put(140,27.5){\vector(2,-1){35}}
\put(140,-5){\vector(1,1){15}}
\put(175,10){\vector(-1,0){20}}
\put(190,-5){\vector(-1,1){15}}
\put(155,10){\vector(2,1){35}}
\put(133,-10){$a$}
\put(192,-10){$b$}
\put(133,30){$c$}
\put(192,30){$d$}
\put(164,3){$e$}
\put(210,5){$= \ \ \ 0$}
\end{picture}
\caption{\footnotesize
STU relation, which is often considered as a pictorial image
of Jacobi identities (\ref{JI}), if we add indices to arrows.
Alternatively, without indices, it serves as a basic constraint in the theory of diagrams,
used in Vogel's $\Lambda$-algebra
and in Kontsevich description of Vassiliev knot invariants.
For simple Lie algebras, where raising of indices is unambiguous, arrows can be eliminated.
}
\end{figure}
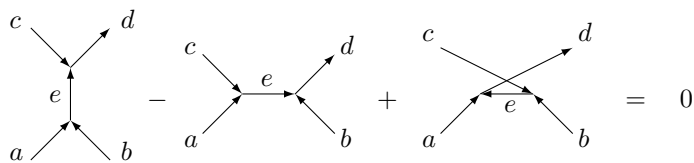

The prominent and constantly increasing
role of symmetries in modern theory \cite{UFN2,UFN3}
calls for better understanding and refreshing of their adequate definition
and classification.
The main role is still played by Lie algebras \cite{Lie} and their minor extensions
to superalgebras \cite{superKac}, affine algebras \cite{KacMoody}, quantum deformations \cite{QA}, Yangians \cite{Yang} and DIM algebras \cite{DIM}.
The concept of Lie algebra is based on the commutator
\be
[T^a, T^b] = f^{ab}_c T^c
\label{LB}
\ee
which satisfies the set of Jacobi identities (JI):
\be
f^{ab}_e f^{ce}_d  + f^{ca}_e f^{be}_d  + f^{bc}_e f^{ae}_d = 0
\label{JI}
\ee
which also has a pictorial realization, see Figure \ref{STUfig}.

Generalizations of Lie algebras sometimes require modification of the Lie bracket (\ref{LB})
and associated deformation/generalization  of JI (\ref{JI}).
Of separate importance is relation of (\ref{JI}) to quadratic Pl\"{u}cker relations, like
\be
W_{12}W_{34} -W_{13}W_{24} + W_{14}W_{23} = 0
\ee
for embedding of Grassmannian $Gr(2,V)$ and its generalization to infinite-dimensional
{\it universal Grassmannian} which have direct relation to integrability theory \cite{UFN3}. As far as we know, Sergei Duzhin was the first to note that Pl\"{u}cker relations lead to weight systems \cite{Du}.\footnote{We thank Sergei Chmutov for pointing these references to us.}
We postpone these part of discussion for the future.

The first question in the theory of Lie algebras {\it per se} is to enumerate the possible structure constants,
i.e. classify the solutions of (\ref{JI}).
This will be the subject of the present review.
The main point is that this problem was solved long ago in the form of the celebrated Dynkin diagrams \cite{Dynkin,superKac}.
Despite this, the problem continued to attract attention, in particular, it led to the no less celebrated Vogel theory \cite{Vogelalg}
-- which, despite widespread expectation, led to a result, very similar to Dynkin's.
{\bf The question is why at all one could expect the difference, and what is the current understanding
and implications of the still remaining discrepancy between Dynkin's and Vogel's results.
A separate story is what are the implications for knot theory, and whether there
exist (Chern-Simons-induced \cite{CS}) knot theories beyond the "ordinary" Lie algebras.}
Surprisingly or not, these are still controversial issues -- reflecting the difficulties of classification
problem for solutions of quadratic relations like (\ref{JI})
and the lack of reliable approaches to this kind of problems.
Our presentation is in no way conclusive and   is rather targeted more at advertising the puzzles rather then
suggesting reliable solutions to them.
Moreover, we do not even pretend to cleverly overview particular subjects like Dynkin, Vogel, Vassiliev or Kontsevich theories,
and can express too naive judgments about them and put improper accents on what is truly essential.
We hope that the story will attract new attention and better structured reviews will appear in the foreseeable future.

\subsection*{Plan of the paper}

We begin in Section \ref{Dynkin} from reminding the Cartan-Dynkin-Kac
classification theory of simple Lie (super)algebras
with emphasize on the peculiar choices, important for the difference with Vogel's approach --
which will be reminded briefly in Sections \ref{Repth} (phenomenology) and \ref{Vogel} (theory).
We omit from the latter section some of the details, already overviewed in our recent \cite{KhMS,KhLS}.

Then we proceed to knot polynomials, which historically were the origin of Vogel's theory.
The main question is to what extent the knot invariants are related/exhausted by the ones, implied by Chern-Simons theory (CST),
which, like and Yang-Mills theory, is essentially based on Lie algebras and thus is hiddenly underlied by JI.
In two different gauges the Yang-Mills CST is reduced to Gaussian, with ultralocal propagators,
selecting either intersection {\it points} in 2d link diagrams (temporal gauge)
or horizontal {\it segments}, connecting intersection points between the link and horizontal planes (holomorphic gauge).
The first choice motivates building knot invariants by Reshetikhin-Turaev theory \cite{RT},
with products of quantum R-matrices inserted at intersection {\it points}.
The second choice motivates writing down a Kontsevich integral \cite{KI, ChD, DBSS} over positions of horizontal sections.
In the second case one can easily obtain an expression with values in (chord) diagrams,
where the role of JI and Dynkin-Vogel restrictions
can look obscure,
while $R$-matrices in the first case are hardly available without/beyond Cartan-Dynkin-Kac-Drinfeld theory.
This causes the long-standing discussion of whether {\it all} Vassiliev invariants -- usually defined as depending on diagrams --
are extractable from R-matrix-induced knot polynomials.
We will provide some details about this discussion in Section \ref{CS}.
Again no final conclusion is made.

Finally in Section \ref{beyond}
we touch the question of going beyond Lie algebras.
It is well known that generalization to {\it quantum} ``groups" is straightforward and widely used in application
to the ordinary knot polynomials.
We briefly remind this story of $q$-deformation of Vogel theory in Section \ref{quant}.
Far more obscure is still the fate of Vogel universality and universal knot polynomials
after the further Macdonald extension(s), which we discuss in Section \ref{sec:upols}.
It is a very interesting question, because on one hand these are related to Yangian and DIM deformations,
and on the other hand they can lead to knot hyper- and/or super-polynomials.
These latter ones are in turn connected to
Khovanov-Rozansky  theory \cite{KhR}, associated with Poincar\'e polynomials of {\it complexes}
rather than their Euler characteristics, reproducing the ordinary knot polynomials.
The subject is even more speculative and controversial than the previous ones,
we briefly comment on it in \ref{Mdknots}.

Section \ref{conc} briefly summarizes our discussion in this paper and
at the Dubna workshop
and outlines some big open directions for the future research in this field.

\section{Dynkin classification
\label{Dynkin}
}

\begin{figure}[h!]
\begin{picture}(850,20)(-25,-5)
\put(30,10){$\textbf{Classical}$}

\put(180,10){$\textbf{Affine}$}

\put(355,10){$\textbf{Super}$}
\end{picture}

\begin{picture}(850,50)(-25,-30)
\put(0,0){\circle{10}}
\put(5,0){\line(1,0){10}}
\put(20,0){\circle{10}}
\put(25,0){\line(1,0){5}}
\put(35,0){\line(1,0){5}}
\put(45,0){\line(1,0){5}}
\put(55,0){\circle{10}}
\put(60,0){\line(1,0){10}}
\put(75,0){\circle{10}}
\put(30,10){$A_n$}

\put(150,0){\circle{10}}
\put(155,0){\line(1,0){10}}
\put(170,0){\circle{10}}
\put(175,0){\line(1,0){5}}
\put(185,0){\line(1,0){5}}
\put(195,0){\line(1,0){5}}
\put(205,0){\circle{10}}
\put(210,0){\line(1,0){10}}
\put(225,0){\circle{10}}
\put(187.5,-20){\circle{10}}
\put(187.5,-20){\color{green}\circle*{9.5}}
\put(153.5,-3.5){\line(2,-1){29.5}}
\put(192,-18.5){\line(2,1){29.5}}
\put(180,10){$\tilde{A}_n$}

\put(300,0){\circle{10}}
\put(305,0){\line(1,0){10}}
\put(320,0){\circle{10}}
\put(325,0){\line(1,0){5}}
\put(335,0){\line(1,0){5}}
\put(345,0){\line(1,0){5}}
\put(355,0){\circle{10}}
\put(360,0){\line(1,0){10}}
\put(375,0){\circle{10}}
\put(380,0){\line(1,0){5}}
\put(390,0){\line(1,0){5}}
\put(400,0){\line(1,0){5}}
\put(410,0){\circle{10}}
\put(371,-2.5){$\times$}
\put(355,10){$\textbf{A}_{m,n}$}
\put(435,6){\small $s=m+1$}
\put(435,-9){\small$r=m+n+1$}
\end{picture}

\begin{picture}(850,60)(-25,-20)
\put(0,0){\circle{10}}
\put(5,0){\line(1,0){10}}
\put(20,0){\circle{10}}
\put(25,0){\line(1,0){5}}
\put(35,0){\line(1,0){5}}
\put(45,0){\line(1,0){5}}
\put(55,0){\circle{10}}
\put(60,0){\line(1,0){10}}
\put(75,0){\circle{10}}
\put(80,1){\line(1,0){15}}
\put(80,-1){\line(1,0){15}}
\put(84,-2.5){$>$}
\put(100,0){\circle{10}}
\put(45,10){$B_n$}

\put(150,10){\circle{10}}
\put(150,-10){\circle{10}}
\put(150,-10){\color{green}\circle*{9.5}}
\put(154.5,8){\line(2,-1){11}}
\put(154.5,-8){\line(2,1){11}}
\put(170,0){\circle{10}}
\put(175,0){\line(1,0){5}}
\put(185,0){\line(1,0){5}}
\put(195,0){\line(1,0){5}}
\put(205,0){\circle{10}}
\put(210,0){\line(1,0){10}}
\put(225,0){\circle{10}}
\put(230,1){\line(1,0){15}}
\put(230,-1){\line(1,0){15}}
\put(234,-2.5){$>$}
\put(250,0){\circle{10}}
\put(195,10){$\tilde{B}_n$}

\put(300,20){\circle{10}}
\put(305,20){\line(1,0){10}}
\put(320,20){\circle{10}}
\put(325,20){\line(1,0){5}}
\put(335,20){\line(1,0){5}}
\put(345,20){\line(1,0){5}}
\put(355,20){\circle{10}}
\put(360,20){\line(1,0){10}}
\put(375,20){\circle{10}}
\put(380,20){\line(1,0){5}}
\put(390,20){\line(1,0){5}}
\put(400,20){\line(1,0){5}}
\put(410,20){\circle{10}}
\put(415,21){\line(1,0){15}}
\put(415,19){\line(1,0){15}}
\put(435,20){\circle{10}}
\put(419,17.5){$>$}
\put(371,17.5){$\times$}
\put(355,30){$\textbf{B}_{m,n}$}
\put(455,26){\small $s=n$}
\put(455,11){\small$r=m+n$}

\put(300,-20){\circle{10}}
\put(305,-20){\line(1,0){10}}
\put(320,-20){\circle{10}}
\put(325,-20){\line(1,0){5}}
\put(335,-20){\line(1,0){5}}
\put(345,-20){\line(1,0){5}}
\put(355,-20){\circle{10}}
\put(360,-20){\line(1,0){10}}
\put(375,-20){\circle{10}}
\put(380,-19){\line(1,0){15}}
\put(380,-21){\line(1,0){15}}
\put(400,-20){\circle*{10}}
\put(384,-22.5){$>$}
\put(355,-10){$\textbf{B}_{0,n}$}
\put(425,-15){\small $s=n$}
\put(425,-30){\small$r=n$}
\end{picture}

\begin{picture}(850,60)(-25,-20)
\put(0,0){\circle{10}}
\put(5,0){\line(1,0){10}}
\put(20,0){\circle{10}}
\put(25,0){\line(1,0){5}}
\put(35,0){\line(1,0){5}}
\put(45,0){\line(1,0){5}}
\put(55,0){\circle{10}}
\put(60,0){\line(1,0){10}}
\put(75,0){\circle{10}}
\put(80,1){\line(1,0){15}}
\put(80,-1){\line(1,0){15}}
\put(84,-2.5){$<$}
\put(100,0){\circle{10}}
\put(45,10){$C_n$}

\put(150,0){\circle{10}}
\put(150,0){\color{green}\circle*{9.5}}
\put(155,1){\line(1,0){15}}
\put(155,-1){\line(1,0){15}}
\put(159,-2.5){$>$}
\put(175,0){\circle{10}}
\put(180,0){\line(1,0){5}}
\put(190,0){\line(1,0){5}}
\put(200,0){\line(1,0){5}}
\put(210,0){\circle{10}}
\put(215,0){\line(1,0){10}}
\put(230,0){\circle{10}}
\put(235,1){\line(1,0){15}}
\put(235,-1){\line(1,0){15}}
\put(239,-2.5){$<$}
\put(255,0){\circle{10}}
\put(200,10){$\tilde{C}_n$}

\put(300,0){\circle{10}}
\put(305,0){\line(1,0){10}}
\put(320,0){\circle{10}}
\put(325,0){\line(1,0){5}}
\put(335,0){\line(1,0){5}}
\put(345,0){\line(1,0){5}}
\put(355,0){\circle{10}}
\put(360,0){\line(1,0){10}}
\put(375,0){\circle{10}}
\put(380,1){\line(1,0){15}}
\put(380,-1){\line(1,0){15}}
\put(400,0){\circle{10}}
\put(385,-2.5){$<$}
\put(296,-2.5){$\times$}
\put(345,10){$\textbf{C}_{n}$}
\put(425,6){\small $s=1$}
\put(425,-9){\small$r=n$}
\end{picture}

\begin{picture}(850,60)(-25,-20)
\put(0,0){\circle{10}}
\put(5,0){\line(1,0){10}}
\put(20,0){\circle{10}}
\put(25,0){\line(1,0){5}}
\put(35,0){\line(1,0){5}}
\put(45,0){\line(1,0){5}}
\put(55,0){\circle{10}}
\put(59.5,2.5){\line(2,1){11}}
\put(59.5,-2.5){\line(2,-1){11}}
\put(75,10){\circle{10}}
\put(75,-10){\circle{10}}
\put(30,10){$D_n$}

\put(150,10){\circle{10}}
\put(150,-10){\circle{10}}
\put(150,-10){\color{green}\circle*{9.5}}
\put(154.5,-8){\line(2,1){11}}
\put(154.5,8){\line(2,-1){11}}
\put(170,0){\circle{10}}
\put(175,0){\line(1,0){5}}
\put(185,0){\line(1,0){5}}
\put(195,0){\line(1,0){5}}
\put(205,0){\circle{10}}
\put(209.5,2.5){\line(2,1){11}}
\put(209.5,-2.5){\line(2,-1){11}}
\put(225,10){\circle{10}}
\put(225,-10){\circle{10}}
\put(180,10){$\tilde{D}_n$}

\put(300,0){\circle{10}}
\put(305,0){\line(1,0){10}}
\put(320,0){\circle{10}}
\put(325,0){\line(1,0){5}}
\put(335,0){\line(1,0){5}}
\put(345,0){\line(1,0){5}}
\put(355,0){\circle{10}}
\put(360,0){\line(1,0){10}}
\put(375,0){\circle{10}}
\put(380,0){\line(1,0){5}}
\put(390,0){\line(1,0){5}}
\put(400,0){\line(1,0){5}}
\put(410,0){\circle{10}}
\put(414.5,2.5){\line(2,1){11}}
\put(414.5,-2.5){\line(2,-1){11}}
\put(430,10){\circle{10}}
\put(430,-10){\circle{10}}
\put(371,-2.5){$\times$}
\put(345,10){$\textbf{D}_{m,n}$}
\put(455,6){\small $s=n$}
\put(455,-9){\small$r=m+n$}
\end{picture}

\begin{picture}(850,60)(-25,-20)
\put(0,0){\circle{10}}
\put(5,0){\line(1,0){10}}
\put(20,0){\circle{10}}
\put(25,0){\line(1,0){10}}
\put(40,0){\circle{10}}
\put(45,0){\line(1,0){10}}
\put(60,0){\circle{10}}
\put(65,0){\line(1,0){10}}
\put(80,0){\circle{10}}
\put(40,5){\line(0,1){10}}
\put(40,20){\circle{10}}
\put(10,10){$E_6$}

\put(150,0){\circle{10}}
\put(155,0){\line(1,0){10}}
\put(170,0){\circle{10}}
\put(175,0){\line(1,0){10}}
\put(190,0){\circle{10}}
\put(195,0){\line(1,0){10}}
\put(210,0){\circle{10}}
\put(215,0){\line(1,0){10}}
\put(230,0){\circle{10}}
\put(190,5){\line(0,1){10}}
\put(190,20){\circle{10}}
\put(190,25){\line(0,1){10}}
\put(190,40){\circle{10}}
\put(190,40){\color{green}\circle*{9.5}}
\put(160,10){$\tilde{E}_6$}
\end{picture}

\begin{picture}(850,60)(-25,-20)
\put(0,0){\circle{10}}
\put(5,0){\line(1,0){10}}
\put(20,0){\circle{10}}
\put(25,0){\line(1,0){10}}
\put(40,0){\circle{10}}
\put(45,0){\line(1,0){10}}
\put(60,0){\circle{10}}
\put(65,0){\line(1,0){10}}
\put(80,0){\circle{10}}
\put(85,0){\line(1,0){10}}
\put(100,0){\circle{10}}
\put(40,5){\line(0,1){10}}
\put(40,20){\circle{10}}
\put(10,10){$E_7$}

\put(150,0){\circle{10}}
\put(150,0){\color{green}\circle*{9.5}}
\put(155,0){\line(1,0){10}}
\put(170,0){\circle{10}}
\put(175,0){\line(1,0){10}}
\put(190,0){\circle{10}}
\put(195,0){\line(1,0){10}}
\put(210,0){\circle{10}}
\put(215,0){\line(1,0){10}}
\put(230,0){\circle{10}}
\put(235,0){\line(1,0){10}}
\put(250,0){\circle{10}}
\put(255,0){\line(1,0){10}}
\put(270,0){\circle{10}}
\put(210,5){\line(0,1){10}}
\put(210,20){\circle{10}}
\put(180,10){$\tilde{E}_7$}
\end{picture}

\begin{picture}(850,60)(-25,-20)
\put(0,0){\circle{10}}
\put(5,0){\line(1,0){10}}
\put(20,0){\circle{10}}
\put(25,0){\line(1,0){10}}
\put(40,0){\circle{10}}
\put(45,0){\line(1,0){10}}
\put(60,0){\circle{10}}
\put(65,0){\line(1,0){10}}
\put(80,0){\circle{10}}
\put(85,0){\line(1,0){10}}
\put(100,0){\circle{10}}
\put(105,0){\line(1,0){10}}
\put(120,0){\circle{10}}
\put(40,5){\line(0,1){10}}
\put(40,20){\circle{10}}
\put(10,10){$E_8$}

\put(150,0){\circle{10}}
\put(155,0){\line(1,0){10}}
\put(170,0){\circle{10}}
\put(175,0){\line(1,0){10}}
\put(190,0){\circle{10}}
\put(195,0){\line(1,0){10}}
\put(210,0){\circle{10}}
\put(215,0){\line(1,0){10}}
\put(230,0){\circle{10}}
\put(235,0){\line(1,0){10}}
\put(250,0){\circle{10}}
\put(255,0){\line(1,0){10}}
\put(270,0){\circle{10}}
\put(275,0){\line(1,0){10}}
\put(290,0){\circle{10}}
\put(290,0){\color{green}\circle*{9.5}}
\put(190,5){\line(0,1){10}}
\put(190,20){\circle{10}}
\put(160,10){$\tilde{E}_8$}
\end{picture}

\begin{picture}(850,60)(-25,-20)
\put(0,0){\circle{10}}
\put(5,0){\line(1,0){10}}
\put(20,0){\circle{10}}
\put(25,1){\line(1,0){15}}
\put(25,-1){\line(1,0){15}}
\put(29,-2.5){$>$}
\put(45,0){\circle{10}}
\put(50,0){\line(1,0){10}}
\put(65,0){\circle{10}}
\put(28,10){$F_4$}

\put(150,0){\circle{10}}
\put(150,0){\color{green}\circle*{9.5}}
\put(155,0){\line(1,0){10}}
\put(170,0){\circle{10}}
\put(175,0){\line(1,0){10}}
\put(190,0){\circle{10}}
\put(195,1){\line(1,0){15}}
\put(195,-1){\line(1,0){15}}
\put(199,-2.5){$>$}
\put(215,0){\circle{10}}
\put(220,0){\line(1,0){10}}
\put(235,0){\circle{10}}
\put(187,10){$\tilde{F}_4$}

\put(320,0){\circle{10}}
\put(325,0){\line(1,0){10}}
\put(340,0){\circle{10}}
\put(345,1){\line(1,0){15}}
\put(345,-1){\line(1,0){15}}
\put(365,0){\circle{10}}
\put(370,0){\line(1,0){10}}
\put(385,0){\circle{10}}
\put(316.5,-2.5){$\times$}
\put(349,-2.5){$<$}
\put(350,10){$\textbf{F}_{4}$}
\end{picture}

\begin{picture}(850,60)(-25,-20)
\put(0,0){\circle{10}}
\put(5,2){\line(1,0){15}}
\put(5,0){\line(1,0){15}}
\put(5,-2){\line(1,0){15}}
\put(25,0){\circle{10}}
\put(10,-2.5){$>$}
\put(8,10){$G_2$}

\put(150,0){\circle{10}}
\put(150,0){\color{green}\circle*{9.5}}
\put(155,0){\line(1,0){10}}
\put(170,0){\circle{10}}
\put(175,2){\line(1,0){15}}
\put(175,0){\line(1,0){15}}
\put(175,-2){\line(1,0){15}}
\put(195,0){\circle{10}}
\put(179,-2.5){$>$}
\put(175,10){$\tilde{G}_2$}

\put(320,0){\circle{10}}
\put(325,0){\line(1,0){10}}
\put(340,0){\circle{10}}
\put(345,2){\line(1,0){15}}
\put(345,-0){\line(1,0){15}}
\put(345,-2){\line(1,0){15}}
\put(365,0){\circle{10}}
\put(316.5,-2.5){$\times$}
\put(349,-2.5){$<$}
\put(340,10){$\textbf{G}_{3}$}
\end{picture}

\begin{picture}(850,60)(-25,-20)
\put(320,0){\circle{10}}
\put(325,0){\line(1,0){10}}
\put(340,0){\circle{10}}
\put(345,0){\line(1,0){10}}
\put(360,0){\circle{10}}
\put(336.5,-2.5){$\times$}
\put(325,10){$\textbf{D}_{2,1}(\lambda)$}
\end{picture}
\vspace{-9mm}\caption{\footnotesize
Dynkin diagrams.
}
\label{Dynkindiag}
\end{figure}
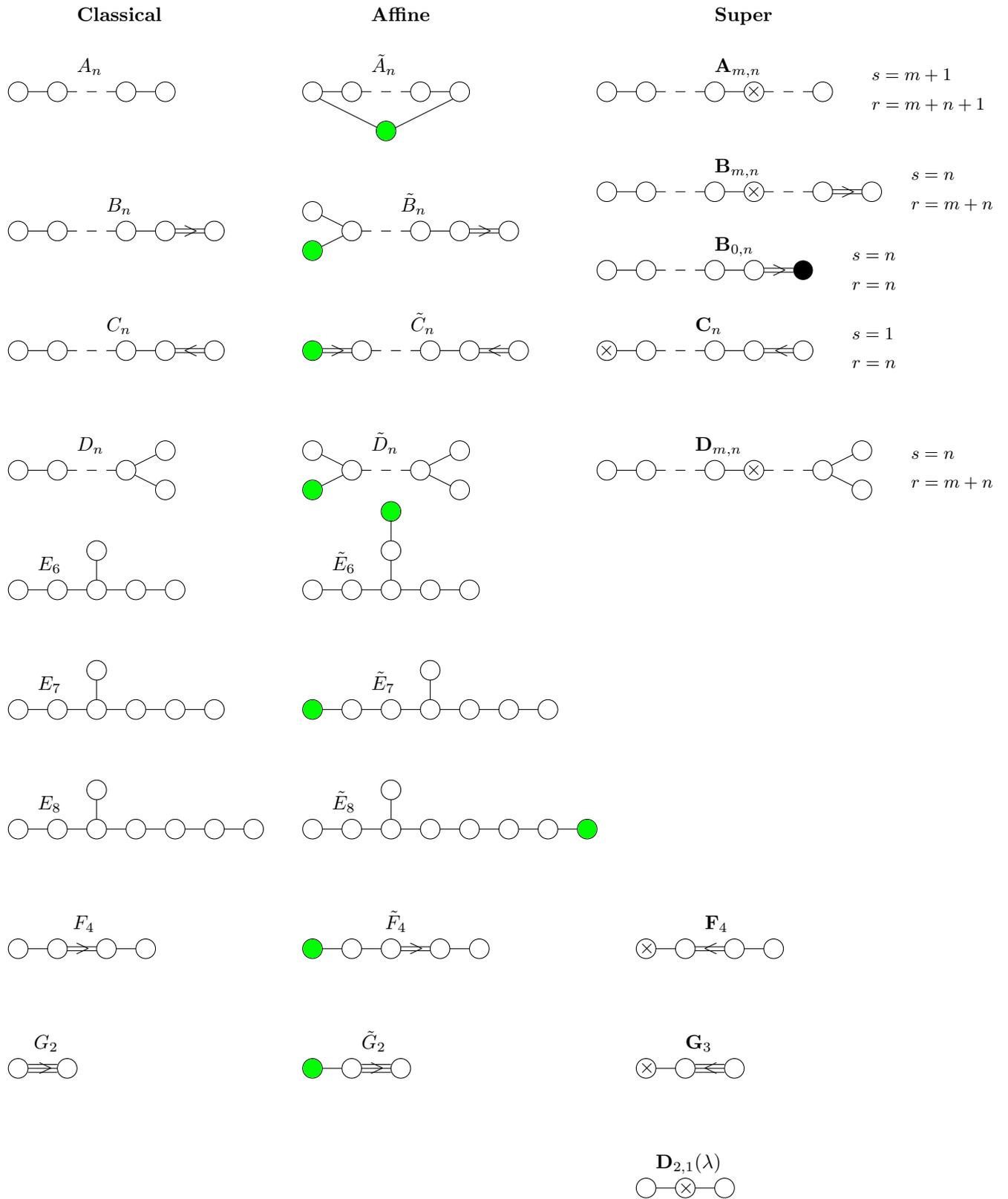

First of all, we remind the list of Dynkin diagrams \cite{Lie, Dynkin} and its most popular extensions, see Figure \ref{Dynkindiag}. We include in this list the cases of affine \cite{KacMoody} and superalgebras \cite{superKac}.
In the first case the finite-growth condition is slightly weakened,
and in the second case some generators are Grassmannian and commutators
are lifted to supercommutators. In the first two columns $n$ is the number of white circles, additional circle for affine case is colored by green. In the last column the index $s$ indicates the position of the only nonwhite circle and the $r$ means the number of the circles. Note that the Dynkin diagrams for the super case no longer uniquely determine the algebra. We give the simplest version of the diagrams as an example.

This first and the third columns represent Dynkin locus on Vogel's plane,
see Figure \ref{figVogelplane} below. Superalgebras add just a single new line $D_{2,1}(\lambda)$ to it.

{\bf Not shown} in the picture are other important classes of non-Dynkin algebras,
which already have important applications in physics:
continuation of $E$-series to $E_9$, $E_{10}$ and $E_{11}$ \cite{West} (perhaps, also further),
further relaxation of Lie structure, allowing "quantum groups",
Yangians and DIM algebras -- in the last two case Dynkin diagrams
are lifted to quiver diagrams, which are still not well classified.

\bigskip

\bigskip

We do not repeat the long reasoning, leading to this list from original Lie and Cartan considerations, instead we just {\it enumerate} the points which have obvious relation to comparison with Vogel's approach:

\begin{itemize}

\item{}\textit{Non-solvable algebras} 

\item{}\textit{Restriction to indices (discreteness)}
\item{}\textit{Restriction to finite set of indices ("finite growth" condition, usually imposed through Serre ids)}

\begin{itemize}
\item{} JI (\ref{JI}) can easily make sense beyond {\it this} finiteness condition, if the sum over $e$ is finite
for given $a,b,c,d$.
This is the case for Kac-Moody (affine) algebras, what leads to a slight extension of Dynkin's set --
though already beyond Vogel theory(!).
In some sense this is true also for Yangians and DIM algebras -- but there the (\ref{JI}) itself should be slightly modified.

\item{} Finiteness is important for traces (diagrams with loops).
\item{} At best it will be substituted by convergence requirements.

\end{itemize}

\item{} \textit{Restriction to simple Lie algebras or to algebras with a unique non-degenerate invariant bilinear form}

\item{} \textit{Extension to superalgebras}

\item{} \textit{Quantization}

\item{} \textit{Extension to affine (Kac-Moody) (super)algebras}

\item{} \textit{Macdonald deformation and extensions towards Yangians and DIM}

\item{} \textit{Multi-parametric metrics}

\end{itemize}

\noindent
The very first and the last three items in this list are the only ones, where
{\it Vogel's universality  in the adjoint sector} still remains a question.
We will not say much about them and concentrate instead on the other points --
where universality exists and discuss the subtleties and open problems
inside this set of theories.

\section{Representation theory and universality in adjoint (``$E_8$'') sector
\label{Repth}
}

An important part of Cartan theory of Lie algebras considers different {\it representations}
of a given algebra, specified by the equivalence class of structure constants $f^{ab}_c$.
The main part of the story is within Cartan approach.
Universality is not obligatory the feature of Vogel theory
(though originally discovered within it, it {\it could} be observed much earlier).
Still some aspects are seen only at the diagram level and are special for Vogel approach.

A separate interesting story is the possibility to build the theory of Lie algebras
in inverse way:
from representation theory rather than from Jacobi identities.
This approach,  known to some as Tannaka-Krein theory \cite{TK} acquires new twists
from the points of view of universality
-- but this is beyond the scope of this short review.

\subsection{Ordinary (Cartan) representation theory}

\begin{itemize}

\item{}
Eq.(\ref{JI}) implies that the matrix $(T^a)^b_c = f^{ab}_c$ satisfies (\ref{LB}).
This is known as {\it adjoint} representation of the Lie algebra.

\item{}
Actually there are many more representations $R$ by matrices ${(T_R^a)}^\mu_\nu$ of different sizes $d_R$, $\mu,\nu=1,\ldots,d_R$.
Moreover, if the two sets of matrices $T_R$ and $T_{R'}$ satisfy (\ref{LB}), then
\be
T_R\otimes I_{d_{R'}} + I_{d_R}\otimes T_{R'} \ \ = \ \oplus_{R''} N_{R,R'}^{R''} T_{R''}
\label{comul}
\ee
with $I_d$ a unit matrix of size $d$, also does so.
The diagrammatic version of this {\it comultiplication rule} is shown in Fig.\ref{comulfig}.
It is not {\it irreducible} and can be further decomposed in irreducible representations (irreps) at the r.h.s. of (\ref{comul}).

\begin{figure}[h]
\begin{picture}(850,80)(-100,-40)
\put(20,-14){\line(-1,6){6}}
\put(40,-14){\line(1,6){6}}
\put(130,-14){\line(2,3){24}}
\put(150,-14){\line(-2,3){24}}
\put(250,-14){\line(0,1){20}}
\put(250,6){\line(1,1){16}}
\put(250,6){\line(-1,1){16}}
\put(80,0){\mbox{\fontsize{16}{16}$-$}}
\put(190,0){\mbox{\fontsize{16}{16}$=$}}
\linethickness{0.6mm}
\qbezier(0,0)(30,-30)(60,0)
\qbezier(110,0)(140,-30)(170,0)
\qbezier(220,0)(250,-30)(280,0)
\put(60,0){\vector(1,1){5}}
\put(170,0){\vector(1,1){5}}
\put(280,0){\vector(1,1){5}}
\put(128,-35){\mbox{\fontsize{13}{13}$\textbf{STU}$}}
\end{picture}
\caption{\footnotesize Diagrammatic version of the comultiplication rule.
Representation $R$ is denoted by thick line, only for $R=adj$ it becomes the ordinary thin line.
${(T^a_R)}^\mu_\nu$ are triple vertices with two thick and one thin tails.
Arrows can be omitted for simple algebras with unambiguous non-degenerate (Killing)  metric. }
\label{comulfig}
\end{figure}
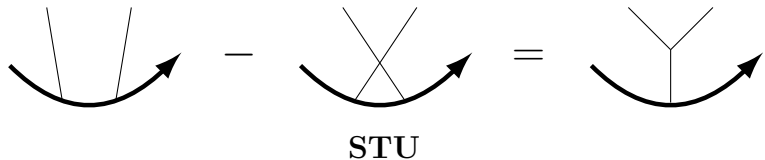

\end{itemize}

\subsection{Universal formulas for adjoint dimensions
\cite{Vogelalg, Deligne,CohenMan,Vogeluniv,LM,RMD,MkrMM,mkrtchyan2012casimir,Mkr,Avet,IP,uniformg4,Mkr_Cas}
\label{unidims}}

For a physicist, the starting observation of universality is that many relevant quantities like
dimensions of various representations, Casimir eigenvalues or Racah matrices can be described by
one and the same (universal) formula for all simple algebras from the Dynkin list --
where different algebras correspond to different choices of the three Vogel parameters $\alpha$, $\beta$, $\gamma$.
Moreover, these formulas can be straightforwardly quantized (possess deformation to quantum groups).

This observation seems true for adjoint sector, i.e. for representations which arise in decompositions of
products of adjoint one -- and sometime correspond not to irreducible representations, but to their combinations,
not separable by knot polynomials.

It is a question, what else in these representations is also universal -- say, entire Schur polynomials
and Littlewood-Richardson coefficients, describing decompositions of representation products.
Also questionable is the next-after-quantum, say, Macdonald deformation or its various reductions, say, to
Jack, Uglov or Hall-Littlewood polynomials.
Moreover, even the claim about universalization of high adjoint dimensions remains to be validated.
In this paper we review some of the existing phenomenology, beginning from the basics in the present section.

\subsubsection{Universal parameters}

The story begins from the spectacular observation that dimensions of all Dynkin algebras,
or of all their adjoint representations are described by a formula
\be
D_{adj} = \frac{(\alpha-2t)(\beta-2t)(\gamma -2t)}{\alpha \beta\gamma}
= \frac{(\hat \alpha-1)(\hat\beta-1)(\hat\gamma -1)}{\hat\alpha \hat\beta\hat\gamma}
\label{Dadj}
\ee
which is a {\it symmetric} rational function of three parameters $\alpha,\ \beta,\ \gamma$,
with $t:=\alpha+ \beta+ \gamma$ and
\be
\hat\alpha:=\frac{\alpha}{2t},\ \hat\beta:=\frac{\beta}{2t},\ \hat\gamma:=\frac{\gamma}{2t}
\ee

The quantities $\alpha,\ \beta,\ \gamma$ for simple Lie algebras were introduced by Pierre Vogel in \cite{Vogelalg}

\be \label{tab:V}
\begin{array}{|c|c|c|c|}
\hline
\text{algebra} & \alpha & \beta & \gamma  \\
\hline
\mathfrak{sl}_{N} &    -2 & 2 & N \\
\hline
\mathfrak{so}_{N} & -2 & 4 & N-4 \\
\hline
\mathfrak{sp}_{2N} & -2 & 1 & N+2\\
\hline
Exc(N) & -2 & N+4 & 2N+4 \\
\hline
\end{array}
\label{VogelT}
\ee
where Vogel's  line of exceptional algebras is
\be
\begin{array}{|c||c|c|c|c|c|c|c|c|c|c|c|c|c|}
\hline &&&&&&&&\\
N & -\frac{4}{3} & -1  & -\frac{2}{3} &0 & 1 & 2 & 4 & 8 \\
&&&&&&&&\\
\hline
Exc(N) & A_1 & A_2 & G_2& D_4 & F_4 & E_6& E_7 & E_8 \\
\hline
\end{array}
\label{VogelExc}
\ee

P.Vogel proposed to unify these algebras into a single series, giving the following definition of exceptional algebras. A simple Lie algebra is called exceptional if the center of its universal enveloping algebra in the 4-th degree is generated by the square of a quadratic Casimir element. Indeed, independent higher Casimir algebras for simple Lie algebras appear in the following orders:

\be
\begin{array}{|c|c|}
\hline &\\
\text{Lie algebra} & \text{Degrees of independent Casimirs} \\
\hline
\mathfrak{sl}_{N+1} & 2, 3, \ldots, N  \\
\hline
\mathfrak{so}_{2N+1} & 2, 4, \ldots, 2N \\
\hline
\mathfrak{so}_{2N} & N; 2, 4, \ldots, 2N-2 \\
\hline
\mathfrak{sp}_{2N} & 2, 4, \ldots, 2N \\
\hline
G_2 & 2, 6 \\
\hline
F_4 & 2,6,8,12 \\
\hline
E_6 & 2,5,6,8,9,12 \\
\hline
E_7 & 2,6,8,10,12,14,18 \\
\hline
E_8 & 2,8,12,14,18,20,24,30 \\
\hline
\end{array}
\label{casimirs}
\ee

From this perspective, exceptional algebras are not as exceptional as the classical series, because they also form a series that, for some profound reason, terminates. We will return to this issue a little further down in  Subsections \ref{vogelplane} and \ref{deligneconj}.

However exceptional algebras are not necessarily organized in a separate line. Alternatively they can be listed one by one:
as shown in Table \ref{tab:char} in the same normalization as in \cite{mkrtchyan2012casimir}:

\be
	\begin{tabular}{|c|c|c|c|c|}
		\hline
		Lie algebra $\mathfrak{g}$& $\alpha$ & $\beta$ & $\gamma$ & $t=\alpha+\beta+\gamma$ \\
		\hline
		$\mathfrak{sl}_{N}$ & $-2$ & $2$ &$N$ &$N$ \\
		\hline
		$\mathfrak{so}_{N}$ & $-2$ & $4$ & $N-4$  &$N-2$\\
		\hline
		$\mathfrak{sp}_{2N}$ & $-2$ & $1$ &$N+2$  &$N+1$\\
		\hline\hline
		$G_2$ & $-2$ & $10/3$ & $8/3$ & $ 4 $ \\
		\hline
		$F_4$ & $-2$ & $5$ &$6$ & $9$ \\
		\hline
		$E_6$ & $-2$ & $6$ & $8$  & $12$\\
		\hline
		$E_7$ & $-2$ & $8$ & $12$ & $18$\\
		\hline
		$E_8$ & $-2$ & $12$ & $20$ & $30$\\
		\hline
	\end{tabular}
\label{tab:char}
\ee

There is no particular difference between (\ref{VogelT}) and (\ref{tab:char}) yet, but as we will see in a moment the choice between  them is somewhat conceptual and can have consequences.

\subsubsection{ Dynkin locus $P_{12}$ and $P_{19}$ from separate exceptional loci }

Symmetric polynomials of $\alpha,\beta,\gamma$ which vanish for particular sets of algebras in  (\ref{VogelT}) are:
\be
\label{Palgebra0}
P_{\mathfrak{sl}} &=& (\alpha+\beta)\,(\beta+\gamma)\,(\alpha+\gamma), \hspace{0mm} \nonumber \\
P_{\mathfrak{osp}} &=& (\alpha+2\beta)(2\alpha+\beta)\,(\beta+2\gamma)(2\beta+\gamma)\,(\alpha+2\gamma)(2\alpha+\gamma),   \\
P_{\mathfrak{exc}} &=& (\alpha-2\beta-2\gamma)\,(\beta-2\alpha-2\gamma)\,(\gamma-2\alpha-2\beta). \nonumber
\ee
Note that symmetrization in $\alpha$,$\beta$,$\gamma$ naturally unite $P_{\mathfrak{so}}$ and $P_{\mathfrak{sp}}$ into a common entity,
in Vogel theory they are inseparable.

The  product
\be
\boxed{
P_{Lie}:= P_{\mathfrak{sl}}\cdot P_{\mathfrak{osp}} \cdot P_{\mathfrak{exc}}
}
\label{Palgebra}
\ee
is the polynomial of  12-th degree, which vanishes on entire  Dynkin list.

But also beyond it: on entire {\it lines} in the $\alpha,\beta,\gamma$-space, which contain a lot of non-integer points.
Moreover, exceptional line contains many points which are integer, but have nothing to do with Dynkin's exceptional algebras.

There is, however, a way to cure at least this latter problem -- if one thinks that it is a problem.
One could make the exceptional factor more precise, by specifying to particular exceptional algebras:
\be
P_{G_2} = 18(\alpha^2+\beta^2+\gamma^2)-25(\alpha+\beta+\gamma)^2 \nn\\
P_{F_4} = 81(\alpha^2+\beta^2+\gamma^2)-65(\alpha+\beta+\gamma)^2 \nn\\
P_{E_6} = 18(\alpha^2+\beta^2+\gamma^2)-13(\alpha+\beta+\gamma)^2 \nn\\
P_{E_7} = 81(\alpha^2+\beta^2+\gamma^2)-53(\alpha+\beta+\gamma)^2 \nn\\
P_{E_8} = 225(\alpha^2+\beta^2+\gamma^2)-137(\alpha+\beta+\gamma)^2
\label{altern_exc}
\ee
The polynomial
\be
\boxed{
{\cal P}_{Lie} = P_{\mathfrak{sl}}\cdot P_{\mathfrak{osp}} \cdot
P_{G_2}\cdot
P_{F_4}\cdot
P_{E_6}\cdot
P_{E_7}\cdot
P_{E_8}
}
\ee
has higher degree $19$, but no longer vanishes at entire exceptional line.

Note that polynomials \eqref{altern_exc} and $P_{\mathfrak{exc}}$ form a ring over symmetrized polynomials $\mathbb{Q}[\alpha, \beta, \gamma]$, each element of which determine the corresponding exceptional Lie algebra. For example, $P_{F_4} = (8\alpha+2\beta+\gamma)\times$ 5 permutations. But not only this increases the degree, such procedure introduces a lot of new integer points along 6 new straight lines. Moreover, the choice of such multi-linear polynomial is highly ambiguous -- what makes the whole idea senseless. Instead, the non-linear choice which we made instead makes the line, which is not straight, still it has just the 2nd order and, even more important, do not pass
through any other integer points besides $F_4$ {\it per se}.

In the next section \ref{Vogel} we will see that in the true Vogel theory the $sl_2$ algebra should be treated more carefully
and this increases the degrees of two polynomials $P_{Lie}$ and ${\cal P}_{Lie}$ by three: to $P_{15}$ and $P_{22}$,
which contain an extra factor
\be
 \omega := (\alpha+\beta+2\gamma)(\beta+\gamma+2\alpha)(\gamma+\alpha+2\beta)
\ee
Additionally they can be multiplied by the contribution of supergroup $D_{2,1}(\lambda)$ which is just
$t=\alpha+\beta+\gamma$,
thus providing
\be
t\cdot P_{15}:= t\cdot \omega\cdot P_{Lie}\ \ \ \ {\rm and} \ \ \ \  t\cdot P_{22}:= t\cdot \omega\cdot {\cal P}_{Lie}
\ee

\subsubsection{Superalgebras}

Vogel's concept of universality also extends to superalgebras. Instead of ordinary dimension, one must use superdimension ($\text{sdim} = \text{dim}_{even} - \text{dim}_{odd}$), instead of trace one must use supertrace, etc \cite{superKac}. This leads to the fact that all simple Lie superalgebras are naturally included in the above considerations, except $D_{2,1}(\lambda)$. For example, Lie superalgebras $\mathfrak{sl}_{m,n}$ have Vogel's parameter $(-2,2,m-n)$. However there are $(9|8)$-dimensional deformations of the $D_{2,1} = \mathfrak{osp}_{4,2}$ depending on parameter $\lambda$. If $\lambda \neq 0,-1$ then these algebras are simple. Its explicit description is given in Section 3 in \cite{Lieberum}.

\subsubsection{Vogel's plane}
\label{vogelplane}
The equations $P_{Lie} = 0$ (see \eqref{Palgebra}) that distinguish Lie algebras are symmetric with respect to the group of permutations of three elements $(\alpha, \beta, \gamma)$ and with respect to the general rescaling. Therefore, these equations can be represented graphically on the plane $\mathbb{P}^2/S_3$, which is called the Vogel's plane.

For example, $sl_N$ series is determined by $P_{sl}=0$ what is equal to $(\alpha+\beta)\,(\beta+\gamma)\,(\alpha+\gamma) = 0$. On the Vogel's plane this series is determined by the line $\beta=2$, because up to permutations we have $\alpha+\beta = 0$ and we fix $\alpha=-2$ by the rescaling according to \eqref{tab:char}. The superalgebra series $D_{2,1}(\lambda)$ is distinguished by the equation $\alpha+\beta+\gamma = 0$ what gives the line $\gamma = 2 - \beta$ on the Vogel's plane. Thus, we get the following picture:

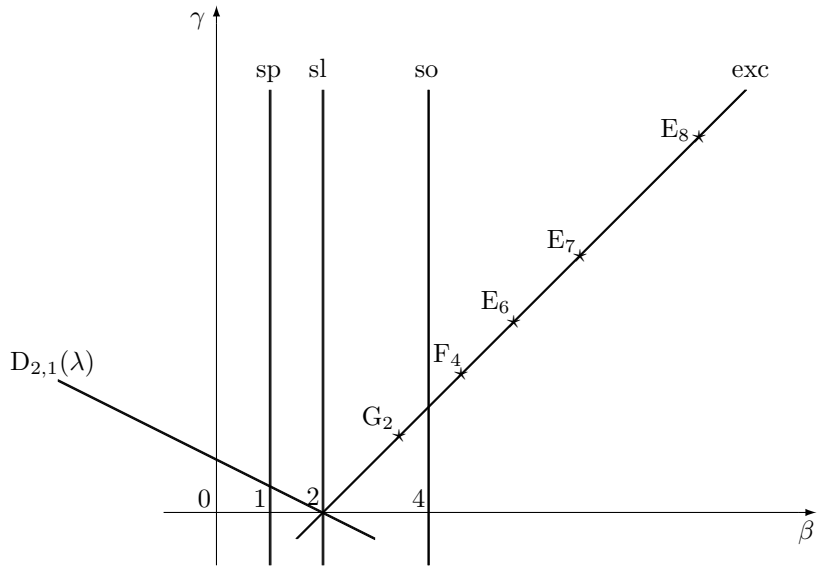
\begin{figure}[H]
\begin{picture}(850,210)(-120,0)
\put(20,0){\line(0,1){205}}
\put(20,205){\vector(0,1){7}}
\put(10,205){$\gamma$}
\put(0,20){\line(1,0){240}}
\put(240,20){\vector(1,0){7}}
\put(240,10){$\beta$}
\put(13,22){$0$}
\put(40,0){\line(0,1){180}}
\put(40.3,0){\line(0,1){180}}
\put(40.6	,0){\line(0,1){180}}
\put(35,185){$\text{sp}$}
\put(34,22){$1$}
\put(60,0){\line(0,1){180}}
\put(60.3,0){\line(0,1){180}}
\put(60.6,0){\line(0,1){180}}
\put(55,185){$\text{sl}$}
\put(54,23){$2$}
\put(100,0){\line(0,1){180}}
\put(100.2,0){\line(0,1){180}}
\put(100.5,0){\line(0,1){180}}
\put(95,185){$\text{so}$}
\put(94,22){$4$}
\put(50,10){\line(1,1){170}}
\put(50.3,10){\line(1,1){170}}
\put(50.6,10){\line(1,1){170}}
\put(215,185){$\text{exc}$}
\put(86.6,46.6){$\star$}
\put(75,53){$\text{G}_2$}
\put(110,70){$\star$}
\put(102,77){$\text{F}_4$}
\put(130,90){$\star$}
\put(120,97){$\text{E}_6$}
\put(155,115){$\star$}
\put(145,120){$\text{E}_7$}
\put(200,160){$\star$}
\put(188,162){$\text{E}_8$}
\put(-40,70){\line(2,-1){120}}
\put(-40,70.3){\line(2,-1){120}}
\put(-40,69.7){\line(2,-1){120}}
\put(-58,74){$\text{D}_{2,1}(\lambda)$}
\end{picture}
\caption{\footnotesize The Dynkin locus on Vogel's plane.
Superalgebras add just a single new line $D_{2,1}(\lambda)$}
\label{figVogelplane}
\end{figure}

If the single exceptional straight line is substituted by individual quadratic factors, \eqref{altern_exc},
then the exceptional locus on Vogel plane takes the following form:

\begin{figure}[H]
	\vspace{0mm}
	\centering\leavevmode
	\includegraphics[width=6.7cm]{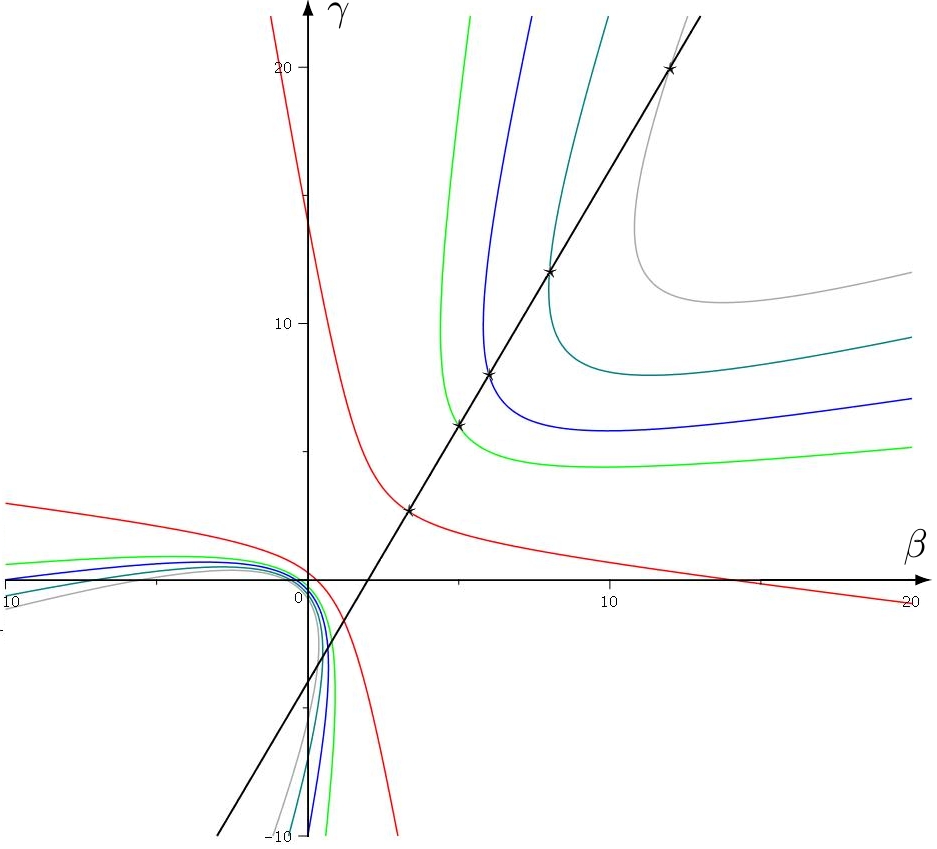}
	\caption{\footnotesize Exceptional locus with linear and quadratic equations. We left Deligne straight line on this picture as well.}
	\vspace{0mm}
\end{figure}

So, we've plotted all the simple Lie algebras on the Vogel plane. The question arises: do the remaining points on this plane correspond to anything meaningful? Let's split this question into two parts.

First, it's clear that all the algebras lie on one of five lines: $\mathfrak{sl}, \ \mathfrak{so}, \ \mathfrak{sp}, \ \mathfrak{exc}, \ D_{2,1}(\lambda)$. The simple (super) Lie algebras are individual points on these lines (actually all algebras have integer points, except for the $G_2$ algebra, whose Vogel parameters are rational in the chosen normalization). The question is whether it's possible to assign meaning to the remaining points on these lines and whether they correspond to any algebras. Second, a similar question can be asked about all the remaining points on the plane that don't belong to these lines.

R. Mkrtchyan discusses this issue in detail \cite{Mkr2}. By analyzing the formula for the quantum dimension \eqref{qdAdj} and requiring it to be no poles, he discovered many new solutions that include both new lines and new isolated points (see also \cite{LM}). The connection with Diophantine equations,  Freudenthal magic squares, and regular polyhedra (Platonic solids) is also discussed. Of course, much remains unclear, especially the algebraic structure, in these new solutions.

The situation is much better with points on the lines $\mathfrak{sl}, \ \mathfrak{so}, \ \mathfrak{sp}, \ \mathfrak{exc}, \ D_{2,1}(\lambda)$. First, it is known that it is possible to assign meaning to the classical series of Lie algebras with complex rank and their representations \cite{Et1, Et2, HarKal}. Second, it was possible to understand the algebraic meaning of the point $(-2,10,16)$. This point was named the algebra $E_{7\frac{1}{2}}$. It is a subalgebra of $E_8$ and contains the algebra $E_7$. Although it is not a simple algebra ($E_{7\frac{1}{2}} = E_7 \, \oplus \, (56) \, \oplus \, \mathbb{R}$), it behaves in many ways like a simple exceptional algebra, for example, it satisfies decomposition and dimension formulas. For details, we refer to \cite{LM75}. In this regard, it is important to mention \textbf{Deligne's conjecture} \cite{Deligne}, which implies that all points on the exceptional line must have a reasonable algebraic meaning.

\subsubsection{Deligne's conjecture}
\label{deligneconj}
We have already seen that the tensor products of the adjoint representation decompose into irreducible representations in a uniform way, and the formulas for the dimensions and values of the Casimir operator are rational functions of the three Vogel parameters.
This is true at least up to the 4-th tensor power \cite{CohenMan}.

Deligne's conjecture \cite{Deligne} is that this will be true for any $k$-th power if we restrict ourselves to the exceptional line $\gamma = 2(\alpha+\beta)$. In this case, each point of this line (not just the 5 exceptional algebras) will have the meaning of some virtual exceptional Lie algebra.

\subsubsection{Currently available list of  dimensions in adjoint sector
}
\label{sec:dim}

Dimension of adjoint representation is a symmetric function of  $\alpha,\beta,\gamma$.
\be
\boxed{
{\rm dim}_{\rm Adj} = \frac{(\alpha-2t)(\beta-2t)(\gamma-2t)}{\alpha\beta\gamma}
}
\label{Dadj}
\ee
where we use the standard notation $t=\alpha+\beta+\gamma$.

\subsubsection*{Square of the adjoint}

Symmetric square of adjoint representation is in general a sum of four representations
(not obligatory irreducible for particular concrete algebras):
\be
{\cal S}^2{\rm Adj} = 1 + Y_2(\alpha)+Y_2(\beta)+ Y_2(\gamma)
\ee
Dimensions of $Y_2$ are no longer symmetric in $\alpha,\beta,\gamma$.
Instead
\be
{\rm dim}_{Y_2(\alpha)} = -\frac{(3\alpha-2t)(\beta-2t)(\gamma-2t) t (\beta+t)(\gamma+t)}{\alpha^2\beta\gamma(\alpha-\beta)(\alpha-\gamma)}
\ee
Dimensions of $Y_2(\beta)$ and $Y_2(\gamma)$ are obtained by permutations.
For exceptional line $3\gamma=2t$, i.e. the factor $3\gamma-2t=0$, what means that representation $Y_2(\gamma)$ disappears
(has zero dimension, to compare, dimension of the singlet is 1).

Antisymmetric square is decomposed into a sum of two, one is again adjoint:
\be
\Lambda^2{\rm Adj} = {\rm Adj} + X_2
\ee
As a consequence, dimension of $X_2$ is again symmetric in $\alpha,\beta,\gamma$:
\be
{\rm dim}_{X_2} = \frac{{\rm dim}_{\rm Adj}({\rm dim}_{\rm Adj}-3)}{2} =
\frac{(2t-\alpha)(2t-\beta)(2t-\gamma)(t+\alpha)(t+\beta)(t+\gamma)}{\alpha^2\beta^2\gamma^2}
\ee

\subsubsection*{Cube of the adjoint}

This time there will be three types of symmetries, labeled by three Young diagrams of size three:
symmetric $([3])$, antisymmetric $([1,1,1]$) and mixed $([2,1])$.

Already in the antisymmetric sector we get a decomposition into  six:
\be
\Lambda^3{\rm Adj} = 1 + X_2 + X_3 +Y_2(\alpha)+Y_2(\beta)+Y_2(\gamma)
\ee
where the only new item $X_3$ is irreducible for exceptional algebras,
but further decomposes into two different irreps for $sl_N$ and $so_N$.
Its dimension is still symmetric in $\alpha,\beta,\gamma$:
\be
{\rm dim}_{X_3} = \frac{{\rm dim}_{\rm Adj}({\rm dim}_{\rm Adj}-1) ({\rm dim}_{\rm Adj}-8)}{6}
\ee
but no longer decomposes into factors, linear in $\alpha,\beta,\gamma$.
It {\it is} decomposed in this way on the exceptional line
\be
\left.{\rm dim}_{X_3}\right|_{\rm Exc} = \frac{10(3N + 8)(N + 1)(5N + 12)(2N + 3)(3N + 7)(5N + 8)}{(N + 4)^3}
\ee
and the r.h.s. is integer at the points $N=-1,-\frac{2}{3},0,1,2,4,8$ at the special points,
associated with the simple exceptional algebras.
For the $sl_N$ and $so_N$ it is instead a sum of two factorized expressions, corresponding to two irreducible representations:
{\footnotesize
\be
\left.{\rm dim}_{X_3}\right|_{sl_N} = \frac{(N + 1)(N - 1)\underline{(N^2 - 2)}(N - 3)(N + 3)}{6}
= \frac{(N^2 - 1)^2(N^2 - 9)}{9}  + \frac{(N^2-1)(N^2-4)(N^2-9)}{18}
\nn \\
\left.{\rm dim}_{X_3}\right|_{so_N} = \frac{((N - 1)N(N - 2)(N + 1)\underline{(N^2 - N - 16)}}{48}
= \frac{N(N^2 - 16)(N - 3)(N^2 - 1)}{72} + \frac{N^2(N^2 - 1)(N + 2)(N - 5)}{144}
\nn
\ee
}
\noindent
Non-factorisable quantities are underlined.

We refer for more details on the third and higher powers of adjoint to \cite{IP, uniformg4}.
These authors used a technique of split Casimir operators, for its pictorial realization in the theory of $\Lambda$-algebras,
to be reviewed in sec.\ref{Vogel} below, see \cite{KhLS}.
Not all irreps are carefully divided by Vogel theory -- typically this is true only for those separated by {\it knot invariants}
\cite{MkrMM},
which were named {\it uirreps} (universal irreps) in \cite{BiMM}.
Not all of emerging dimensions, already at the forth power, are deduced in \cite{IP, uniformg4} in the nice factorized form --
what is, in particular an obstacle for their quantization on the lines of sec.\ref{quant}.
Remarkably, the problematic quantities are {\it not} contributing to torus {\it knots}, at least for 4 strands --
i.e. for ${\rm Adj}^{\otimes 4}$ \cite{BiMi2}.
Significance and generality of this fact remain unclear at this moment.

\subsection{Moral}

In this section we described the simple but impressive phenomenology behind universality in the adjoint sector
of algebras from the Dynkin list.

As with every phenomenology, a number of questions immediately arises, concerning the {\it ad hoc} choices,
which seem somewhat arbitrary and could be changed.
A partial list of these choices include
\begin{itemize}
\item{} polynomiality in $\alpha,\beta,\gamma$,
\item{} symmetry in $\alpha,\beta,\gamma$,
\item{} minimality -- like the choice of $P_{12}$ against its multiplication by any other function
\end{itemize}

As we already mentioned, polynomiality is broken by  quantum deformation(s).

Symmetry looks like a rather arbitrary feature of (\ref{Dadj}), and, as we saw, permutations of $\alpha$, $\beta$, $\gamma$
act non-trivial already on the dimensions of irreps in $Adj^{\otimes 2}$.

A next step is a more  careful theory about contractions of structure constants, restricted by Jacobi identities (\ref{JI}),
to which we proceed in the next section \ref{Vogel}.
In particular we will learn something about the minimality issue -- $P_{12}$ will indeed be extended to $P_{15}=\omega\cdot P_{12}$.

\section{Vogel theory
\label{Vogel}
}

\subsection{Pictures instead of formulas}
Consider a vector space $\Lambda$ over $\mathbb{Q}$ generated by 3-legged (the legs are numbered) diagrams with 1- and 3-valent vertices modulo AS \eqref{LB} and IHX \eqref{JI} relations antisymmetric with respect to permutations of legs:
\begin{figure}[H]
\begin{picture}(850,80)(-60,-30)
\put(0,0){\line(0,1){20}}
\qbezier(0,20)(10,25)(-10,45)
\qbezier(0,20)(-10,25)(10,45)
\put(20,20){$= \ \ \ -$}
\put(60,0){\line(0,1){20}}
\put(60,20){\line(-1,2){12.5}}
\put(60,20){\line(1,2){12.5}}
\put(20,-20){\mbox{\fontsize{13}{13}$\textbf{AS}$}}

\put(170,10){\line(0,1){25}}
\put(170,35){\line(-2,1){20}}
\put(170,35){\line(2,1){20}}
\put(170,10){\line(-2,-1){20}}
\put(170,10){\line(2,-1){20}}
\put(210,20){$=$}
\put(238,44){\line(1,-2){12}}
\put(238,-4){\line(1,2){12}}
\put(250,20){\line(1,0){25}}
\put(287,44){\line(-1,-2){12}}
\put(287,-4){\line(-1,2){12}}
\put(305,20){$-$}
\put(330,0){\line(1,2){10}}
\put(330,43.5){\line(3,-2){35}}
\put(340,20){\line(1,0){25}}
\put(375,0){\line(-1,2){10}}
\put(340,20){\line(3,2){35}}
\put(250,-20){\mbox{\fontsize{13}{13}$\textbf{IHX}$}}
\end{picture}

\caption{AS and IHX relations are diagrammatic presentation of Lie bracket and Jacobi identity}
\end{figure}
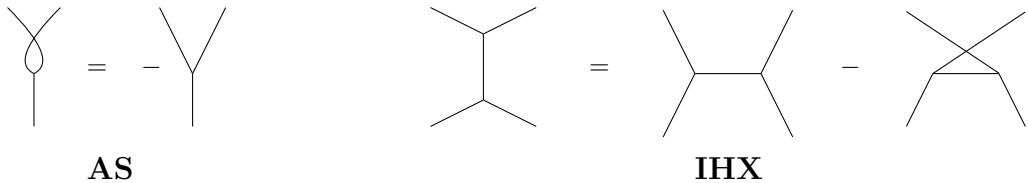

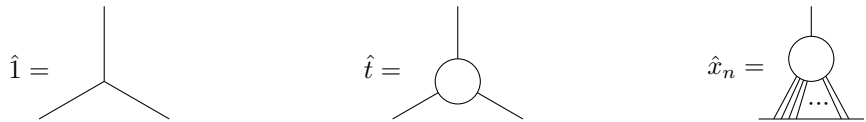
\begin{figure}[!h]
	\centering
	\begin{tikzpicture}
		
		\node at (-5.7, 0.2) {$\hat{1}=$};
		\draw (-4.7,0)--+(90:1);
		\draw (-4.7,0)--+(-30:1);
		\draw (-4.7,0)--+(210:1);
		
		\node at (-1, 0.2) {$\hat{t}=$};
		\draw (90:0.3)--(90:1);
		\draw (-30:0.3)--(-30:1);
		\draw (210:0.3)--(210:1);
		\draw (0,0) circle (0.3);
		
		\node at (3.7, 0.2) {$\hat{x}_n=$};
		\draw (4., -0.5) -- (5.4, -0.5);
		\draw (4.7, 0.6) -- +(0, 0.4);
		\draw (4.7, 0.3) circle (0.3);
		\draw ($(4.7, 0.3)+(50:-0.3)$) -- (4.2, -0.5);
		\draw ($(4.7, 0.3)+(60:-0.3)$) -- (4.3, -0.5);
		\draw ($(4.7, 0.3)+(70:-0.3)$) -- (4.4, -0.5);
		\draw ($(4.7, 0.3)+(80:-0.3)$) -- (4.5, -0.5);
		\node at (4.7, -0.3) {$\cdot$};
		\node at (4.8, -0.3) {$\cdot$};
		\node at (4.9, -0.3) {$\cdot$};
		\draw ($(4.7, 0.3)+(120:-0.3)$) -- (5.1, -0.5);
		\draw ($(4.7, 0.3)+(130:-0.3)$) -- (5.2, -0.5);
	\end{tikzpicture}
	\caption{Some elements of $\Lambda$}
	\label{fig:some-of-Lambda}
\end{figure}

All legs for any 3-valent vertex are equivalent, hence, we {\bf omit arrows} in contrast to Figure \ref{STUfig}.
For Lie algebra this can be done when there is an invariant bilinear form
which eliminates the difference between upper and lower indices and we can switch
to the fully antisymmetric structure constant $f^{abc}$.

On the other hand, we {\bf do not require here discreteness or finiteness} of indices $a,b,c$.
Still we need that the definition of index contraction is unambiguous.
Except for the single vertex, 3-leg diagrams necessarily include loops, i.e. traces,
and their definition beyond finite discrete sums requires finiteness or regularization conditions.

\bigskip

{\bf Still, how much difference could one {\it expect}?
Why could one hope to get anything far beyond Dynkin?}
The answer is not quite clear to us,
and eq.(\ref{tP15}) below does not look like a too big  surprise.

\subsection{$\Lambda$-algebra}

One can promote the vector space $\Lambda$, defined above,
to the algebra with the help of a vertex multiplication: insert one diagram into any vertex of the another diagram. For example,

\begin{picture}(850,80)(-145,-40)
\put(0,0){\circle{20}}
\put(0,10){\line(0,1){15}}
\put(-7,-7){\line(-1,-1){10}}
\put(7,-7){\line(1,-1){10}}
\put(30,0){$\cdot$}
\put(60,0){\circle{20}}
\put(60,10){\line(0,1){15}}
\put(53,-7){\line(-1,-3){4.3}}
\put(60,-10){\line(0,-1){10}}
\put(67,-7){\line(1,-3){4.3}}
\put(40,-20){\line(1,0){40}}
\put(90,0){$=$}
\put(130,10){\circle{10}}
\put(130,15){\line(0,1){10}}
\put(123,-6){\line(-1,-3){4.7}}
\put(130,-10){\line(0,-1){10}}
\put(137,-6){\line(1,-3){4.7}}
\put(110,-20){\line(1,0){40}}
\qbezier(125,10)(110,0)(130,-10)
\qbezier(135,10)(150,0)(130,-10)
\end{picture}

It is well-defined due to AS and IHX relations. One can prove that this multiplication is commutative.

$\Lambda$-algebra is graded: $\Lambda = \oplus_{n=0}^{\infty} \, \Lambda_n,$ where $2n = \#\text{vertices} - 1$. It is obvious that AS and IHX relations do not change the degree $n$, which is additive under the multiplication. 	

In general the structure of {\it commutative and associative graded} $\Lambda$-algebra is unknown.
{\it Hypothetically}, $\hat x_n$ provide a multiplicative basis in it -- which is actually not free,
there are many relations between $\hat x_n$.  At least up to degree 10 the $\Lambda$-algebra is isomorphic to the polynomial algebra in just 3 variables $(t, \sigma, \omega)$,
e.g.
\be
\label{basis}
\hat x_1 &=& 2\hat t, \nn \\
\hat x_3 &=& 4\hat t^3 - \frac{3}{2} \hat \omega,   \\
\hat x_5 &=& 12 \hat t^5 - \frac{17}{2}\hat t^2 \hat \omega + \frac{3}{2} \hat \sigma \hat \omega. \nn
\ee

Lie algebra $\mathfrak{g}$ defines a linear function $\Phi_{\mathfrak{g}}$ on $\Lambda$-algebra associating the structure constant $f^{abc}$ with a vertex
\be
\Phi_{\mathfrak{g}} \Big(
\put(16,3){\line(0,1){12}}
\put(16,3){\line(3,-2){10.5}}
\put(16,3){\line(-3,-2){10.5}}
\put(16,3){\circle*{2}}
\hspace{11mm}\Big) = f^{abc}.
\nonumber
\ee
If a Lie algebra $\mathfrak{g}$ has only one invariant 3-tensor (i.e. only the structure constant $f^{abc}$), then the value of $\Phi_{\mathfrak{g}}$ on any diagram from $\Lambda$-algebra is proportional to the $f^{abc}$:
\be
\Phi_{\mathfrak{g}}(\hat{x}) = \chi_{\mathfrak{g}}(\hat{x}) \, f^{abc}, \quad \forall\, \hat{x} \in \Lambda.
\ee
Coefficient $\chi_{\mathfrak{g}}(\hat{x})$ is a well-defined \textit{character} of $\Lambda$.

\subsection{Universality}
Character $\chi_{\mathfrak{g}}(\hat{x})$ maps any diagram to the polynomial in $t, \sigma$ and $\omega$. Values of these variables depend on the Lie algebra $\mathfrak{g}$. Thus, for the 3-legged 3-valent diagrams we get formulas universal for any $\mathfrak{g}$. For illustrative purpose let us consider some examples:
\be
\chi_{\mathfrak{g}}(\hat x_1) &=& 2 t, \nn \\
\chi_{\mathfrak{g}}(\hat x_3) &=& 4 t^3 - \frac{3}{2} \omega,   \\
\chi_{\mathfrak{g}}(\hat x_5) &=& 12 t^5 - \frac{17}{2} t^2  \omega + \frac{3}{2}  \sigma  \omega.
\ee
Unlike formula \eqref{basis}, here $t$, $\sigma$ and $\omega$ are no longer diagrams, but rather the values of their characters, so there are no hats above them. To find the explicit values of these characters, we must calculate the corresponding contraction of the structure constants of the algebra $\mathfrak{g}$. Since the diagrams $\hat x_1$, $\hat x_3$, and $\hat x_1$ form a multiplicative basis, as discussed in the previous subsection, it is sufficient to calculate the characters for only these diagrams to determine the values of the variables $t$, $\sigma$, and $\omega$.

There is also a convenient parameterization using three parameters $\alpha$, $\beta$ and $\gamma$, which have already appeared in Section \ref{Repth}:
\be
\alpha + \beta +\gamma = t, \quad \alpha \beta + \beta \gamma + \alpha \gamma = \sigma-2t^2, \quad \alpha \beta \gamma = \omega-t\sigma.
\ee
Their explicit values are given in Table \ref{tab:char}.

Also one can obtain universal formulas also for 2-legged and 0-legged 3-valent diagrams. Any 2-legged diagram is proportional to the metric (invariant bilinear form) times 3-legged diagram. Any 1-legged diagram is zero due to AS relation. 0-legged diagrams form the algebra of 3-graphs, which is isomorphic to the $\Lambda$-algebra \cite{ChDK}.

\subsection{Dynkin's implications}

The natural question is to
look for diagrams with characters $P_{15}$ and $P_{22}$.\footnote{
The possible existence of something like $P_{22}$,
which {\it could} break the Deligne hypothesis about
exceptional line, was emphasized to us by R.Mkrtchyan.}
Perhaps, they can be interesting.
At least $P_{15}$ is.

\subsection{Zero divisors}

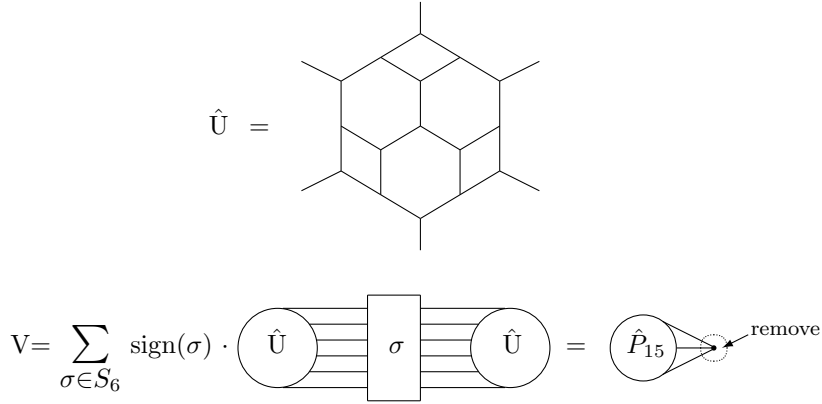
\begin{figure}[H]
\begin{picture}(300,120)(-150,-15)
\put(0,50){$\hat{\text{U}} \ \, = $}

\put(50,53){\line(0,1){17}}
\put(80,53){\line(0,1){17}}
\put(50,70){\line(5,3){15}}
\put(65,79){\line(5,-3){15}}
\put(50,53){\line(5,-3){15}}

\put(110,53){\line(0,1){17}}
\put(80,70){\line(5,3){15}}
\put(95,79){\line(5,-3){15}}
\put(95,44){\line(5,3){15}}

\put(65,27){\line(0,1){17}}
\put(95,27){\line(0,1){17}}
\put(65,44){\line(5,3){15}}
\put(80,53){\line(5,-3){15}}
\put(80,18){\line(5,3){15}}
\put(65,27){\line(5,-3){15}}

\put(50,36){\line(0,1){17}}
\put(50,36){\line(5,-3){15}}

\put(110,36){\line(0,1){17}}
\put(95,27){\line(5,3){15}}

\put(80,88){\line(5,-3){15}}
\put(65,79){\line(5,3){15}}

\put(80,88){\line(0,1){12}}
\put(80,6){\line(0,1){12}}
\put(50,70){\line(-2,1){15}}
\put(50,36){\line(-2,-1){15}}
\put(110,36){\line(2,-1){15}}
\put(110,70){\line(2,1){15}}

\end{picture}

\begin{picture}(850,40)(-375,-45)
\put(-300,-22){V= {\Large \text{$\sum \limits_{\sigma \in S_6}$}} $ \text{sign}(\sigma) \ \cdot$}
\put(-199,-20){\circle{30}}
\put(-203,-22){$\hat{\text{U}}$}

\put(-197,-35){\line(1,0){32}}
\put(-187,-29){\line(1,0){22}}
\put(-184.5,-23){\line(1,0){19.5}}
\put(-184.5,-17){\line(1,0){19.5}}
\put(-187,-11){\line(1,0){22}}
\put(-197,-5){\line(1,0){32}}

\put(-165,-40){\line(0,1){40}}
\put(-165,-40){\line(1,0){20}}
\put(-165,-0){\line(1,0){20}}
\put(-145,-40){\line(0,1){40}}
\put(-157,-22){$\sigma$}

\put(-145,-35){\line(1,0){32}}
\put(-145,-29){\line(1,0){22}}
\put(-145,-23){\line(1,0){19.5}}
\put(-145,-17){\line(1,0){19.5}}
\put(-145,-11){\line(1,0){22}}
\put(-145,-5){\line(1,0){32}}

\put(-111,-20){\circle{30}}
\put(-114,-22){$\hat{\text{U}}$}

\put(-90,-22){$=$}
\put(-60.7,-20){\circle{25}}
\put(-67,-22){$\hat{P}_{15}$}
\put(-48,-20){\line(1,0){14.5}}
\put(-53.3,-10){\line(2,-1){20}}
\put(-53.3,-30){\line(2,1){20}}
\put(-34,-20){\circle*{2}}
\qbezier[8](-34,-15)(-29,-15)(-29,-20)
\qbezier[8](-29,-20)(-29,-25)(-34,-25)
\qbezier[8](-34,-25)(-39,-25)(-39,-20)
\qbezier[8](-39,-20)(-39,-15)(-34,-15)
\put(-20,-15){$\footnotesize \text{remove}$}
\put(-21,-15){\vector(-2,-1){10}}
\end{picture}

\caption{\footnotesize
The image of $P_{15}$ in $\Lambda$-algebra
}
\label{tP15fig}
\end{figure}

It turns out that the $\Lambda$-algebra is not a polynomial algebra, because it has zero divisors.
The simplest one is
\be
\boxed{
\hat t\cdot \hat P_{15}=0
}
\label{tP15}
 \ee
as a consequence of {\it just} JI -- without any reference to Dynkin's analysis (!).
Here
\be
\chi_{\mathfrak{g}}(\hat{P}_{15}) = P_{sl_2}\,P_{sl}\,P_{osp}\,P_{exc}
\ee
Pictorially the statement (\ref{tP15}) is depicted in Figure \ref{tP15fig}. The image of $P_{15}$ in $\Lambda$-algebra is made from linear combinations of diagrams
obtained by different contractions of 6 external lines of two "bubbles" $\hat{\rm{ U}}$ (a sum over permutations). To obtain a 3-legged diagram, which belongs to $\Lambda$-algebra, we erase one of the trivalent vertex -- in fact, any, they are all equivalent modulo JI. After that we can insert the diagram $\hat t$. Miraculously (?), the resulting diagram is identical zero modulo JI.

An obvious way to construct many more zero divisors is given by multiplication  $\hat{t}\cdot \hat{P}_{15} \cdot \hat{x}_n = 0$. Are there more zero divisors beyond this family?
It is still an open question, whether the same is true for $P_{22}$:
\be
\hat t\cdot \hat P_{22} =\ ?
\label{tP22}
\ee
 where
\be
\chi_{L}(\hat{P}_{22}) = P_{sl_2}\,P_{sl}\,P_{osp}\,(77t^2-36\sigma)(176t^2-81\sigma)(494t^2-225\sigma)(170t^2-81\sigma)(65t^2-36\sigma)
\ee
The calculation behind (\ref{tP15}) is quite difficult and additional effort is needed to lift it to more complicated cases
like (\ref{tP22}).

The zero divisor $\hat{t}\cdot \hat{P}_{15} = 0$ means that the character $\chi_{L}$ of any Lie algebra $L$, which determines linear functions $\Phi_{L}$ on $\Lambda$-algebra, must be zero either on $\hat{t}$ or on $\hat{P}_{15}$.
It gives an alternative way to classify Lie algebras,
but still leads to the Cartan-Dynkin-Kac classification of simple (super)algebras.

\be
\boxed{
 \text{ {\bf Is it a surprise -- did not Dynkin prove exactly this?}}
}
\nn
\ee

\subsection{Moral}

Steps which were made:

\begin{itemize}
\item{} Unification of formulas for dimensions/traces/characters
    $\Longrightarrow$ symmetric functions of $\alpha,\beta,\gamma$, integer valued  for Dynkin cases.
   However, there are integer points beyond it, and some of them are already associated with non-Dynkin Lie algebras.
   Moreover, this appears possible only in the adjoint/$E_8$ sector and not in the full representation theory.
   Explanations of these peculiarities are still not very clear.

\item{} Multiplication of diagrams, unambiguous due to (\ref{JI}), gives $\Lambda$-algebra.
But why/what for?
Characters are coefficients and involve traces.
Multiplication in $\Lambda$ is given by insertion of one diagram into any vertex of the other diagram.

\item{} Hypothesis of $\hat x_n$ basis (still not free $\longrightarrow$ 3 parameters).

\item{} { Sectors of tree and loop diagrams.
Strictly speaking, in Vogel algebra of 3-legged diagrams  there is a single tree diagram, which is just the structure constant,
but other multi-legged tree diagrams could also be studied.
What is important,
there are no universal formulas for such diagrams, as well as for structure constants {\it per se} -- only for traces.
This may be not a too big surprise, since structure constants are defined modulo changes of basis and/or Weyl reflections.
}

\item{} Look for diagrams with characters $P_{15}$ and $P_{22}$.
Perhaps, they can be interesting.

\item{} At least $\hat P_{15}$ is:
\be
\boxed{
\hat t\cdot \hat P_{15}=0
}
\label{ttP15}
 \ee
as a consequence of {\it just} JI -- without any reference to Dynkin's analysis (!).
Pictorially this statement is depicted in/represented by
Fig.\ref{tP15fig}.

\item{}
It is still an open question, whether the same is true for $\hat P_{22}$
\be
\hat t\cdot \hat P_{22} =\ ?
\ee

\end{itemize}

\section{Chern-Simons theory and knot polynomials
\label{CS}
}

The 3-dimensional Chern-Simons action with gauge group $G$ for vector field $A_{\mu} = A_{\mu}^a(x)\,T^a$, where $T^a$ are generators of the corresponding Lie algebra $\mathfrak{g}$, is
\be
\label{CSaction}
S(A) = \dfrac{\kappa}{4\pi}\int_{S^3}d^3x \,\epsilon^{\mu\nu\rho}\,\Tr\left(A_{\mu}\partial_{\nu}A_{\rho} + \frac{2}{3}A_{\mu}A_{\nu}A_{\rho}\right).
\ee

Gauge invariant functions of $A_{\mu}$ are Wilson loops
\be\label{wilson}
\langle W_R(\mathcal{K}) \rangle = \dfrac{1}{Z} \int \mathcal{D}A \ {\rm e}^{iS(A)} \ \Tr_R\left( {\rm Pexp} \ i\oint_CA_{\mu}dx^{\mu} \right),
\ee
that determine topological invariants of contour $C$. Partition function is defined by $Z = \int \mathcal{D}A \ {\rm e}^{iS(A)}$. These functions \eqref{wilson} are known as quantum knots invariants or Witten-Reshetikhin-Turaev invariants \cite{CS, RT}. They equal to invariants calculated with the help of quantum R-matrices in the corresponding representation of quantum algebra $U_q(\mathfrak{g})$.

\subsection{Two gauges }
Since the Chern-Simons theory is a gauge theory, then the result of calculation does not depend on the gauge fixing procedure, but details of computations do depend. Moreover, a clever choice of the gauge can reveal interesting hidden structures underlying the correlation functions.
The main simplification is elimination of non-linearity.
For Chern-Simons action this is naturally achieved in  the temporal \cite{temporal} and holomorphic \cite{holomorphic, DBSS} gauges, $A_0=0$ and $\bar A=0$.

\subsubsection{Temporal gauge and RT formalism}

Let us fix $A_0=0$. The action \eqref{CSaction} becomes quadratic
\be
S(A) = \dfrac{\kappa}{4\pi}\int_{S^3}\Tr \left( A_x\dot{A_y} - A_y\dot{A_x} \right)dt\,dx\,dy
\ee
and the propagator is
\be
\label{temp_prop}
\langle\, A_{m}^{a}(\vec{x}), \, A_{n}^{b}(\vec{x}') \,\rangle = \dfrac{\hbar}{2\pi i} \, \epsilon_{m n} \,\delta^{a b}\,  \text{sign}(t-t') \,  \delta^{(2)}(\vec{x}-\vec{x}'), \quad m,n=1,2,
\ee
with $\hbar=\frac{2\pi i}{\kappa}$.

One can consider a Wilson line operator
\be
U(\vec{x}_1,\vec{x}_2) = {\rm Pexp} \left( i\int (A_{x}dx + A_ydy) \right)
\ee
Delta-functional propagator \eqref{temp_prop} implies that only special points on the line contributes, namely, crossing points and turning points:

\begin{picture}(450,100)(-100,-30)
\put(0,-10){\line(0,1){60}}
\put(0.2,-10){\line(0,1){60}}
\put(0.4,-10){\line(0,1){60}}
\put(0.2,47){\vector(0,1){7}}
\put(3,50){$y$}
\put(277,-10.2){\vector(1,0){7}}
\put(280,-7){$x$}
\put(0,-10){\line(1,0){280}}
\put(0,-10.2){\line(1,0){280}}
\put(0,-10.4){\line(1,0){280}}

\put(20,0){\line(1,1){40}}
\put(38,22){\line(-1,1){18}}
\put(60,0){\line(-1,1){18}}
\put(37,45){$\mathcal{R}$}

\qbezier(100,10)(130,50)(160,10)
\put(127,45){$\mathcal{Q}$}
\qbezier(200,30)(230,-10)(260,30)
\put(227,45){$\mathcal{Q}^{-1}$}
\end{picture}

In result the knot invariant is expressed in terms of just
two basic operators $\mathcal{R}$ and $\mathcal{Q}$, which satisfy the following equations:
\be
\label{qtr}
\tr_2\left( \mathcal{R}^{\pm} \ 1\otimes\mathcal{Q} \right) = q^{\pm C_2} \\
\mathcal{R}\, \mathcal{R}^{-1} = \mathbb{I} \\
\left(\mathcal{R}\otimes1\right) \left(1\otimes\mathcal{R}\right) \left(\mathcal{R}\otimes1\right) = \left(1\otimes\mathcal{R}\right) \left(\mathcal{R}\otimes1\right) \left(1\otimes\mathcal{R}\right)
\ee
where $C_2 = \sum_{\alpha \in \triangle} h_{\alpha^{\vee}}h_{\alpha}
+ \sum_{\alpha \in \Phi^{+}} (e_{\alpha}f_{\alpha} + f_{\alpha}e_{\alpha})$
is a quadratic Casimir operator, $\triangle$ is the set of simple roots, $\Phi^{+}$ is the set of positive roots, $\alpha^{\vee}$ is a dual to the root $\alpha$. These equations are the algebraic forms of the three Reidemeister moves, which provide the topological invariant (ambient isotopy) of the line or the contour.

Explicit description of the operator $\mathcal{R}$ can be given in terms of quantized universal enveloping algebras $U_{\hbar}(\mathfrak{g})$ with the help of quantum R-matrix:
\be
\label{Rmatr}
\mathcal{R} = \mathcal{P}\, q^{\sum\limits_{\alpha \in \varDelta}h_{\alpha}\otimes h_{\alpha^{\vee}}}
\prod_{\beta \in \Phi^{+}} \exp_{q_{\beta}}
\Big[( q_{\beta}-q^{-1}_{\beta}) E_\beta\otimes F_\beta\Big]\,,
\ee
where $E_\beta$ and $F_\beta$, $\beta \in \Phi^{+}$, are the root elements of $U_{\hbar}(\mathfrak{g})$, and $\mathcal{P}(x\otimes y) = y\otimes x$ is a permutation operator. The q-exponent is defined as follows:
\be
\exp_{q}(X) = \sum_{n=0}^{\infty} \dfrac{X^n}{[n]_q!} \, q^{n(n	-1)/1}, \\
{[n]_q!} = \prod_{k=1}^{n} [k]_q, \quad [k]_q = \dfrac{q^k-q^{-k}}{q-q^{-1}}, \quad q_{\beta} = q^{(\beta,\beta)/2}.
\ee

The turning point operator is defined as follows:
\be
\mathcal{Q} = q^{h_{\rho}}, \quad {\rho} = \dfrac{1}{2} \sum_{\beta \in \Phi^{+}} \beta, \quad h_{\rho} = \sum_{\beta \in \Phi^{+}} h_{\beta}.
\ee

In order to give a general explicit formula for the Wilson loop correlator \eqref{wilson} in terms of $\mathcal{R}$ and $\mathcal{Q}$ a knot needs to be presented in the corresponding form. For example, one can use a plait presentation or a braid presentation. The Alexander theorem states that any knot or link can be presented as a closure of the corresponding braid. So, let us briefly discuss the braid group representation.

The Artin braid group can be defined by generators and relations as follows:
\be
B_n = \{ \sigma_1, ..., \sigma_{n-1} \big| \ \sigma_i\,\sigma_{i+1}\,\sigma_i = \sigma_{i+1}\,\sigma_i\,\sigma_{i+1}, \ \sigma_i\,\sigma_j = \sigma_j\,\sigma_i, \ \forall\, |i-j|\neq1 \}
\ee
Let $V_i, \ i=1..n,$ are $U_{\hbar}(\mathfrak{g})$-modules. One can define the following operators
\be
\mathcal{R}_i = \mathbb{I}_{V_1}\otimes \ldots \otimes \mathcal{R}_{V_i,V_{i+1}} \otimes \ldots \otimes \mathbb{I}_{V_{n}},
\ee
which acts on $V_1\otimes \ldots \otimes V_{n}$ identically everywhere except $V_i\otimes V_{i+1}$, where it acts by the quantum R-matrix \eqref{Rmatr} taken in the corresponding representation. These operators $\mathcal{R}_1, \ldots, \mathcal{R}_{n-1}$ define a representation of the braid group $B_n$ on $n$ strands:
\be
\pi: B_n &\rightarrow& \text{End}( V_1, \ldots, V_n ), \nn\\
\pi(\sigma_i) &=& \mathcal{R}_i.
\ee

In order to get a knot we must consider a closure of a braid, what corresponds to taking a trace \eqref{qtr}. Thus, the Wilson loop average of the knot $\mathcal{K}$ can be calculated as follows
\be
q^{-w(\mathcal{K}) C_2(R)} \dfrac{\langle\, W^{\mathfrak{g}}_R(\mathcal{K})\,\rangle}{\langle\, W^{\mathfrak{g}}_R(\bigcirc)\,\rangle}= q^{-w(\hat{b}) C_2(R)} \dfrac{\tr_R\Big( \pi(b) \, \mathcal{Q}^{\otimes n} \Big)}{\tr_R\left( \mathcal{Q}^{\otimes n} \right)},
\label{InvRT}
\ee
where such normalization provides a polynomial behavior on $q$. This is known as the Reshetikhin-Turaev (RT) \cite{RT} formalism of building up knot and link polynomials, and it can be developed in many directions.

This {\it universality}, however, is not quite the same as Vogel's universality,
and the question of obtaining the Vogel-universal knot polynomials, expressed through symmetric combinations of parameters
$\alpha,\beta,\gamma$ is still open.
A strong support for existence of such polynomials -- in adjoint/$E_8$ sector -- comes from the two facts \cite{MkrMM}:
\begin{itemize}
\item{} RT polynomials are usually expressed through {\it quantum} dimensions and Casimir eigenvalues,
\item{} these quantities in adjoint/$E_8$ sector have Vogel-universal description, see sec.\ref{quant} below\footnote{
Moreover, there are delicate and impressive details: some representations are not {\it separated} in Vogel theory --
but this happens exactly when they are combined together in formulas for RT knot/links polynomials.
Vogel universality is actually about a {\it part} of representation theory:
the one, which is captured by adjoint knot/link polynomials.
}.
\end{itemize}

Relation between the two universalities -- Drinfeld's
and Vogel's   --
is one of the challenging problems.
It is about quantum groups  and about $R$-matrices.
While quantization ($q$-deformation) is consistent with Vogel theory,
its full-scale promotion to $R$-matrices (in adjoint sector)
is still to be made.
A partial step is the theory of Racah matrices in the simplest representations in adjoint sector,
and it looks doable, see \cite{MMracah}.

\subsubsection{Holomorphic gauge}

Let us fix $A_{\bar{z}}=0$. The action \eqref{CSaction} becomes quadratic
\be
S(A) = i\,\displaystyle\int dt d\bar{z} dz\,\epsilon^{m n} A_{m}^{a} \partial_{\bar{z}} A_{n}^{a}
\ee
and the propagator becomes
\be
\langle\, A_{m}^{a}(t_{1},z_{1},\bar{z}_{1})\, A_{n}^{b}(t_{2},z_{2},\bar{z}_{2}) \,\rangle = \dfrac{\hbar}{2\pi i} \, \epsilon_{m n} \,\delta^{a b}\, \dfrac{\delta(t_{1}-t_{2})}{z_{1}-z_{2}}
\ee
with $\hbar=\frac{2\pi i}{\kappa + N}$. Then according to the Wick theorem we get the following answer
{\small
	\be
	\label{wilsonhol}
	\langle\, W^{\mathfrak{g}}_R(\mathcal{K})\,\rangle=\sum\limits_{n=0}^{\infty} \,\dfrac{\hbar^n}{(2\pi i)^n}\,\displaystyle\int\limits_{o(z_{1})<...<o(z_{n})}\,\sum\limits_{p\,\in P_{2n}}\,(-1)^{p_\downarrow}\,
	\bigwedge\limits_{k=1}^{n}\, \dfrac{dz_{i_k}-dz_{j_k} }{z_{i_k}-z_{j_k}}\cdot {\Tr}_{R}\Big( T^{a_{\sigma_{p}(1)}} ...T^{a_{\sigma_{p}(2n)}} \Big),
	\ee
}
where the integration is over the contour $\mathcal{K}$ with the given orientation $o$; the sum runs over the set of all pairing $P_{2n}$ of $2n$ numbers, an element of this set has the form $p=((i_1,j_1)...(i_n,j_n))$ where $i_k < j_k$ and the numbers $i_k,j_k$ are all different numbers from the set $\{1, 2, . . . , 2n\}$. A function $\sigma_p$ returns a minimum number from the pair $(i_k , j_k)$. A symbol $p_\downarrow$ denotes the number of down-oriented segments between critical points on the knot $\mathcal{K}$ entering the integral. More details one can find in \cite{holomorphic, DBSS}.

\subsection{Perturbative expansion of quantum invariants}

Knot polynomials depend in the quantum parameter $q=e^\hbar$, they become trivial (unities) when $\hbar=0$.
A natural thing to do is to consider the expansion in powers of $\hbar$.
The coefficients of this expansion depend on the knot/link ${\cal K}$ and on the gauge algebra $\mathfrak{g}$. It is possible to separate these dependencies and obtain knot invariants, which are numerical, and group-independent \cite{holomorphic, DBSS}:
\be
\left<W^{\mathfrak{g}}_R(\cal K)\right> = \sum_{n=0}^\infty \hbar^n \sum_{m=1}^{\dim {\mathcal{G}}_n}  G^{\mathfrak{g},R}_{n,m} \,v^{\cal K}_{n,m},
\label{hbarexpansion}
\ee
where $\mathcal{G}_n$ is a number of independent group factors of the order $n$. These $v^{\cal K}_{n,m}$ are the celebrated Vassiliev invariants \cite{Vassiliev}.
They, however, have an a priori different definition, and it is a non-trivial exercise to check that they are equivalent.

A procedure is known which generalizes (\ref{wilsonhol}) to guarantee the equivalence, moreover it can be made consistent with Vogel's universality
(for this one should restrict to representations $R$ from adjoint sector only).
However, a similar generalization of (\ref{InvRT}) is not yet (?) available.

\subsection{Vassiliev invariants}

This is an ultrashort reminder of what the Vassiliev invariants are, for details see \cite{ChD}. First of all, we define a singular knot as an isotopy class of $S^1$ immersions in $\mathbb{R}^3$ such that all self-intersection points are simple double points with transversal intersections.

\noindent Then we define the continuation of knot invariants to singular knots by \textit{Vassiliev skein relation} shown in Fig.\,\ref{fig:Vass-skein}. This is a rule for resolving double points of a singular knot. It is clear that a value of an invariant at a singular knot does not depend on an order in which its double points are resolved.
	\begin{figure}[h]
		\centering
		\begin{tikzpicture}[scale=0.4]
			\coordinate (a) at (-7, 0);
			\coordinate (b) at (-0.5, 0);
			\coordinate (c) at (6, 0);
			\begin{pgfonlayer}{background layer}
				\draw [-{Stealth[length=3mm]}, ultra thick] ($(c)-(1, 1)$) -- ($(c)+(1, 1)$);
				\draw [-{Stealth[length=3mm]}, ultra thick] ($(b)-(-1, 1)$) -- ($(b)+(-1, 1)$);
				\draw [-{Stealth[length=3mm]}, ultra thick] ($(a)-(1, 1)$) -- ($(a)+(1, 1)$);
				\draw [-{Stealth[length=3mm]}, ultra thick] ($(a)-(-1, 1)$) -- ($(a)+(-1, 1)$);
			\end{pgfonlayer}
			\begin{pgfonlayer}{main}
				\draw [white, line width=10pt] ($(c)+(1, -1)$) -- ($(c)+(-1, 1)$);
				\draw [-{Stealth[length=3mm]}, ultra thick] ($(c)+(1, -1)$) -- ($(c)+(-1, 1)$);
				
				\draw [white, line width=10pt] ($(b)+(-1, -1)$) -- ($(b)+(1, 1)$);
				\draw [-{Stealth[length=3mm]}, ultra thick] ($(b)+(-1, -1)$) -- ($(b)+(1, 1)$);
				\draw [fill] (a) circle(0.15);	
			\end{pgfonlayer}
			\begin{pgfonlayer}{foreground layer}
				\draw [dashed] (a) circle (1.4142);
				\draw [dashed] (b) circle (1.4142);
				\draw [dashed] (c) circle (1.4142);
				\node [scale =1.6] at ($(a)+(-2.4, 0)$) { \textit{v} } ;
				\node [scale =2.5] at ($(a)+(-1.7, 0)$) { ( } ;
				\node [scale =2.5] at ($(a)+(1.7, 0)$) { ) } ;
				\node [scale =1.6] at ($(b)+(-2.4, 0)$) { \textit{v} } ;
				\node [scale =2.5] at ($(b)+(-1.7, 0)$) { ( } ;
				\node [scale =2.14] at ($(b)+(1.7, 0)$) { ) } ;
				\node [scale =1.6] at ($(c)+(-2.4, 0)$) { \textit{v} } ;
				\node [scale =2.5] at ($(c)+(-1.7, 0)$) { ( } ;
				\node [scale =2.5] at ($(c)+(1.7, 0)$) { ) } ;
				\node [scale=1.8] at ($0.6*(b) + 0.4*(c)+(0.2, 0)$) {$-$};
				\node [scale=1.8] at ($0.6*(a) + 0.4*(b)+(0.2, -0.05)$) {$=$};
			\end{pgfonlayer}
		\end{tikzpicture}
\vspace{-3mm}
		\caption{Vassiliev skein relation.}
		\label{fig:Vass-skein}
	\end{figure}

\noindent An invariant $v$ is called a Vassiliev invariant of order no more than $n$ if $v$ vanishes at all singular knots with more than $n$ double points. The space of Vassiliev invariants of order $\leq n$ is denoted $\mathcal{V}_n$.

There is a some combinatorial description of Vassiliev invariants in terms of chord diagrams. A chord diagram of order $n$ is an oriented circle with $n$ pairs of distinct points. A linear span of chord diagrams of order $n$ is denoted $\mathbf{A}_n$. Algebra of chord diagrams is a vector space $\mathcal{A}_n = \large\textbf{A}_n/\langle\text{4T, 1T} \rangle$  modulo the four-term and the one-term relations with a well-defined multiplication:
\begin{figure}[H]
	\centering
\begin{tikzpicture}[scale=0.5]
\put(-70,-3){$4T =$}
\coordinate (a) at (-2, 0);
\coordinate (b) at (1, 0);
\coordinate (c) at (4, 0);
\coordinate (d) at (7, 0);
\draw[dashed] (a) circle (1);
\draw[thick] ($(a)+(240:1)$) arc(240:300:1);
\draw[thick] ($(a)+(0:1)$) arc(0:60:1);
\draw[thick] ($(a)+(140:1)$) arc(140:180:1);
\draw ($(a)+(255:1)$)--($(a)+(160:1)$);
\draw ($(a)+(280:1)$)--($(a)+(20:1)$);

\draw[dashed] (b) circle (1);
\draw[thick] ($(b)+(240:1)$) arc(240:300:1);
\draw[thick] ($(b)+(0:1)$) arc(0:60:1);
\draw[thick] ($(b)+(140:1)$) arc(140:180:1);
\draw ($(b)+(285:1)$)--($(b)+(160:1)$);
\draw ($(b)+(255:1)$)--($(b)+(20:1)$);

\draw[dashed] (c) circle (1);
\draw[thick] ($(c)+(240:1)$) arc(240:300:1);
\draw[thick] ($(c)+(0:1)$) arc(0:60:1);
\draw[thick] ($(c)+(140:1)$) arc(140:180:1);
\draw ($(c)+(45:1)$)--($(c)+(160:1)$);
\draw ($(c)+(270:1)$)--($(c)+(20:1)$);

\draw[dashed] (d) circle (1);
\draw[thick] ($(d)+(240:1)$) arc(240:300:1);
\draw[thick] ($(d)+(0:1)$) arc(0:60:1);
\draw[thick] ($(d)+(140:1)$) arc(140:180:1);
\draw ($(d)+(20:1)$)--($(d)+(160:1)$);
\draw ($(d)+(270:1)$)--($(d)+(45:1)$);

\node at ($0.5*(a)+0.5*(b)$) {$-$};
\node at ($0.5*(c)+0.5*(b)$) {$-$};
\node at ($0.5*(c)+0.5*(d)$) {$+$};
\end{tikzpicture} \qquad \qquad
\begin{tikzpicture}[scale=0.5]
\put(47,10){$,$}
\put(57,10){$1T =$}
\draw[dashed] (7, 1) circle (1);
\draw[thick] (7, 1)+(160:1) arc(160:200:1);
\draw[thick] (7, 1)+(340:1) arc(340:380:1);
\draw (6,1)--(8,1);
\draw[dotted] ($(7, 1)+(160:1)$) --($(7, 1)+(20:1)$);
\draw[dotted] ($(7, 1)+(200:1)$) --($(7, 1)+(340:1)$);
\end{tikzpicture}
\end{figure}
\hspace{-5mm}
\be
\begin{picture}(850,5)(-380,-35)
\put(-280,-23){multiplication:}
\put(-178,-20){\circle{32}}
\put(-194,-20){\line(1,0){32}}
\put(-178,-36){\line(0,1){32}}
\put(-158,-22){$\cdot$}
\put(-143,-34){\line(0,1){28}}
\put(-127,-34){\line(0,1){28}}
\put(-135,-36){\line(0,1){32}}
\put(-151,-20){\line(1,0){32}}
\put(-135,-20){\circle{32}}
\put(-108,-22){$=$}
	
\put(-90,-10){\line(1,-3){9}}
\put(-90,-30){\line(1,3){9}}
	
\put(-74,-2){\line(1,-2){14}}
\put(-65,-5){\line(-1,-3){11}}
\put(-59,-12){\line(-1,-3){8}}
\put(-62,-8){\line(-1,-3){10}}
\put(-75,-20){\circle{36}}
\end{picture}
\label{anmult}
\ee
One can check that Vassiliev invariants also satisfy 4T and 1T relations, hence they provide linear functions on the algebra of chord diagrams. Let $\mathcal{W}_n = \mathcal{A}_n^{*} = \text{Hom}~(\mathcal{A}_n, \mathbb{R})$ be a space of linear functions on $\mathcal{A}_n$. Then $ \bigoplus_{n=0}^{\infty} \mathcal{W}_n \cong \bigoplus_{n=0}^{\infty} \mathcal{V}_n/\mathcal{V}_{n-1}$ (see details in \cite{Barnatan,ChD,KhLS}). Thus, Vassiliev invariants form a faithful system of linear functions on $\mathcal{A}_n$.

\subsection{Kontsevich integral vs quantum invariant}

The generalization of Wilson loop correlator \eqref{wilsonhol} is the Kontsevich integral, which takes values in the algebra of chord diagrams:
\be
\label{KI}
\text{KI}(\mathcal{K}) = \sum_{n=0}^{\infty} \hbar^n \sum_{m=1}^{\dim {\mathcal{V}}_n} D_{n,m} \cdot v_{n,m}^{\mathcal{K}},
\ee
where $D_{n,m}$ are basis elements of $\mathcal{A}_n$. We do not discuss integral form \eqref{wilsonhol}
of Vassiliev invariants what is beyond of our scope.

It can seem that the formulas (\ref{hbarexpansion}) and (\ref{KI}) are literally the same, but {\bf this is not so}.
They differ by a change of the algebra-dependent coefficients for chord diagrams.
It is easy to restore (\ref{hbarexpansion}) from (\ref{KI}):
one can define a linear function $\varphi^{\mathfrak{g}}_{R}$ on chord diagrams, which respects 4T relations from the algebra of chord diagrams. Then the group factor ${\Tr}_{R}\Big( T^{a_{\sigma_{p}(1)}} T^{a_{\sigma_{p}(2)}}...T^{a_{\sigma_{p}(2n)}} \Big)$ from Wilson loop \eqref{wilsonhol} equals to $\varphi^{\mathfrak{g}}_{R}( D_{n,m} )$ for the corresponding chord diagram. A chord diagram is equivalent to a trivalent 0-legged diagram, hence, it has a universal formula in the Vogel sense.

Still an inverse lift from (\ref{hbarexpansion}) to (\ref{KI}) is problematic:
some chord diagrams can be annihilated by all the maps $\varphi^{\mathfrak{g}}_{R}$ for all representations --
and then (\ref{KI}) would turn more general than (\ref{hbarexpansion}) and \eqref{wilsonhol}.

The algebra of chord diagrams is isomorphic to the algebra of Vassiliev invariants $v_{n,m}^{\mathcal{K}}$ according to the Vassiliev-Kontsevich theorem \cite{KI}. Therefore, the Kontsevich integral \eqref{KI}, which takes values in the algebra of chord diagrams, contains all Vassiliev invariants --
including the ones associated to diagrams, annihilated by all $\varphi^{\mathfrak{g}}_{R}$.
And such diagrams exist -- $D_{17}$ is the simplest example, constructed by P.Vogel in \cite{Vogelalg}.

\bigskip

A rigorous mathematical presentation of the method for constructing the $D_{17}$ diagram with all the necessary proofs can be found in Sections 7-8 of the original paper \cite{Vogelalg}. A less formal presentation is given in Section 2.3.5 of the recent paper \cite{KhMS}. Therefore, we will not repeat this construction here, but will only briefly describe the idea itself.

The idea is based on the Vogel algebra, its characters and the polynomials of the Lie algebra series. The essence of the idea is to construct such a combination of chord diagrams that the weight system $\varphi^{\mathfrak{g}}_{R}$ on it is proportional to the corresponding polynomial $P_{\mathfrak{g}}$. However, if we try to implement this idea literally, then the desired combination of chord diagrams must be proportional to the product of all Lie algebra polynomials $t\, P_{sl}\, P_{osp}\, P_{exc}\, \omega$. And as we know, no diagram corresponds to such a polynomial due to the presence of zero divisors in the Vogel algebra $\Lambda$. Therefore, we must act a little more carefully.

First, we need to go to the basis of closed Jacobi diagrams, in which the action of the Vogel algebra is arranged in a natural way, similar to the multiplication inside $\Lambda$ itself. The algebra of chord diagrams is isomorphic to the algebra of closed Jacobi diagrams, which is the vector space of closed diagrams with only trivalent $2n$ vertices and distinguished cycle modulo STU relations:
\begin{figure}[H]
	\centering
	\begin{tikzpicture}[scale=0.5]
	\coordinate (s) at (0,0);
	\draw[-{Stealth[length=3mm]}, ultra thick] ($(s)+ (45:-2)$) arc(225:315:2);
	\coordinate (t) at (4,0);
	\draw[-{Stealth[length=3mm]}, ultra thick] ($(t)+ (45:-2)$) arc(225:315:2);
	\coordinate (u) at (8,0);
	\draw[-{Stealth[length=3mm]}, ultra thick] ($(u)+ (45:-2)$) arc(225:315:2);
	\coordinate (s1) at ($(s)+(-1, 0)$);
	\coordinate (t1) at ($(t)+(-1, 0)$);
	\coordinate (u1) at ($(u)+(-1, 0)$);
	\coordinate (s2) at ($(s)+(1, 0)$);
	\coordinate (t2) at ($(t)+(1, 0)$);
	\coordinate (u2) at ($(u)+(1, 0)$);
	\draw ($(s)+(0, -2)$) -- ($(s)+(0, -1)$);
	\draw (s1) -- ($(s)+(0, -1)$);
	\draw ($(s)+(0, -1)$) -- (s2);
	\draw[fill] ($(s)+(0, -1)$) circle(0.1);
	
	\draw (t1) -- ($(t)+(70:-2)$);
	\draw ($(t)+(110:-2)$) -- (t2);
	
	\draw (u1) -- ($(u)+(110:-2)$);
	\draw ($(u)+(70:-2)$) -- (u2);
	\node[scale=1.5] at ($0.5*(s) + 0.5*(t)-(0,1)$) {$-$};
	\node[scale=1.5] at ($0.5*(u) + 0.5*(t)-(0, 1)$) {$+$};
	\node[scale=1.5] at ($0.5*(s) - 0.9*(t)-(0,1)$) {$STU = $};
	\end{tikzpicture}
	\label{fig:STU-relation}
\end{figure}
\noindent
It is easy to see that STU relation implies 4T relation. Also STU relation itself implies AS and IHX relation on the internal graphs of Jacobi diagrams. Therefore, there is a natural action of $\Lambda$-algebra on the internal (connected) graph of closed Jacobi diagram.

Second, we need to consider the diagram $\hat{P}_{15}$, whose character is equal to the polynomial $P_{15}$. This diagram itself is no longer equal to 0 and can be multiplied by the simplest closed Jacobi diagram
\be
\label{weight17}
\hspace{0mm} D_{17} := \hat{P}_{15} \cdot
\put(12,3){\circle*{24}}
\put(12,3){\color{white}\circle*{21}}
\put(12,3){\circle*{2}}
\put(12,3){\line(0,1){12}}
\put(12,3){\line(2,-1){10.5}}
\put(12,3){\line(-2,-1){10.5}} \qquad \ \ .
\ee

It is important that $D_{17} \neq 0.$ Let's calculate the weight system of this diagram:
\be
\label{weight170}
\varphi_{\mathfrak{g}}\left(D_{17}\right) = \varphi_{\mathfrak{g}}\left( \hat{P}_{15} \cdot \hat{t}  \right) \cdot \varphi_{\mathfrak{g}}\Big(
\put(12,3){\circle*{24}}
\put(12,3){\color{white}\circle*{21}}
\put(0,3){\line(1,0){24}}
\put(24,0){$\Big)$}
\put(32,0){$=0.$}
\nonumber
\ee

Thus, this diagram is equal to 0 for any simple Lie algebra $\mathfrak{g}$, and it does not contribute to \eqref{hbarexpansion}. Accordingly, the coefficient in front of this diagram (the corresponding Vassiliev invariant) is absent from the quantum invariants.

$$
\boxed{\text{
\bf How can it be?
}}
$$
$$
\text{JI seems to guarantee Reidemeister invariance.}
$$
$$
\boxed{\text{
If $D_{17}$ vanishes due to JI, how can one get an invariant, associated to it?
}}
$$

Once again, we do not know a solution to this puzzle.
For us it remains one more open question.

\subsection{Moral}
In this Section we have defined the knot invariant using Chern-Simons theory in two ways. The first way is non-perturbative, which gives a quantum knot polynomial. The second way is perturbative, which gives an answer in the form of a series in the Vassiliev invariants. Since both ways are related by a gauge transformation, the invariants are the same, but presented in completely different forms.

Both answers have a universality different from the Vogel universality. In the first case, the answer is expressed in terms of the quantum universal R-matrix. In the second case, the answer can be extended to the Kontsevich integral, the universal Vassiliev invariant.

\

Some open questions:
\begin{itemize}
\item{} Express the universal R-matrix in terms of the universal Vogel parameters.
\item{} Write the universal Vassiliev invariant in the universal Vogel form.
\item{} Find a Lie algebra that detects the $D_{17}$ diagram, or explain why such algebra does not exist.
\item{} How to construct an explicit Vassiliev invariant for $D_{17}$?
\item{} Is it possible to present two such nodes that are not distinguished by any quantum polynomials, but differ by the $D_{17}$ Vassiliev invariant (and therefore differ by the Kontsevich integral).
\end{itemize}

\section{Slightly beyond Lie algebras
\label{beyond}
}

This section is a brief reminder that Vogel theory has a $q$-deformation --
not yet developed from quantum dimensions and {\it some} Racah matrices to {\it universal} $R$-matrices --
but encounters problems in further $q,t$-deformation to Macdonald sector,
which is currently under intensive investigation.

\subsection{Quantization
\label{quant}}

\subsubsection{Quantum deformation of Vogel algebra}

Quantization of adjoint dimensions (deformation to those of quantum group)
imply the change of Vogel parameters to
\be
u = q^\alpha, \ \ v=q^\beta, \ \ w=q^\gamma
\ee
We use the standard notation $\{x\}:=x-\frac{1}{x}$.
Classical dimensions from Section \ref{unidims} are reproduced in the limit $q\rightarrow 1$. The list of universal dimensions begins from
\be
{\cal D}_{Adj} = -\frac{\{\sqrt{u}vw\}\{\sqrt{v}uw\}\{\sqrt{w}uv\}}{\{\sqrt{u}\}\{\sqrt{v}\}\{\sqrt{w}\}}
\label{qdAdj}
\ee
The square of adjoint representation is decomposed into six
(they are not obligatory irreducible, but become such, if simple
algebras are multiplied by the automorphisms of their Dynkin diagrams).
Four of them belong to symmetric square, they enter with plus signs
 into the Rosso-Jones formula,
remaining ones -- into the antisymmetric square and they enter with minuses.
Dimensions are known in the universal form
\cite{MkrMM}:
\be
{\cal S}^2 {\cal D}_{Adj} = \frac{{\cal D}_{Adj}^2(q) +{\cal D}_{Adj}(q^2)}{2}
= 1+D_{Y_2}(\alpha)+D_{Y_2}(\beta)+D_{Y_2}(\gamma) \nn \\
{\Lambda}^2 {\cal D}_{Adj} = {\cal D}_{Adj} +  \frac{{\cal D}_{Adj}^2(q) -{\cal D}_{Adj}(q^2)
-2{\cal D}_{Adj}(q)}{2} = {\cal D}_{Adj} + {\cal D}_{X_2}
\ee
with
\be
{\cal D}_{Adj} = -\frac{\{\sqrt{u}vw\} \{u\sqrt{v}w\}\{uv\sqrt{w}\} }
{\{\sqrt{u}\} \{\sqrt{v}\}\{\sqrt{w}\} } \nn \\
{\cal D}_{Y_2}(\alpha) =
\frac{\{uvw\}\{u\sqrt{v}w\}\{uv\sqrt{w}\}\{v\sqrt{uw}\}\{w\sqrt{uv}\}\{vw/\sqrt{u}\}}
{\{\sqrt{u}\}\{u \}\{\sqrt{v}\}\{\sqrt{w}\}\{\sqrt{u/v}\}\{\sqrt{u/w}\}}
\nn \\
{\cal D}_{X_2} = {\cal D}_{Adj}\cdot\frac{\{u\sqrt{vw}\}\{v\sqrt{uw}\}\{w\sqrt{uv}\}}{\{u\}\{v\}\{w\}}
\left(\sqrt{uv}+\frac{1}{\sqrt{uv}}\right)\left(\sqrt{vw}+\frac{1}{\sqrt{vw}}\right)
\left(\sqrt{uw}+\frac{1}{\sqrt{uw}}\right)
\label{dimssquare}
\ee
The corresponding universal expression for quadratic Casimir operators are \cite{Vogeluniv}:
\be
\lambda_{Adj} = q^{t}= uvw, \ \ \ \ \
\lambda_{Y_2}(\alpha) = q^{2t-\alpha} = uv^2w^2, \ \ \ \ \
\lambda_{X_2} = q^{2t} = (uvw)^2
\label{evssquare}
\ee
We use here notation from \cite{Vogelalg, Vogeluniv} for particular descendants of adjoint representation.
Formulas for $Y_2(\beta)$ and $Y_2(\gamma)$ are obtained from $Y_2(\alpha)$
by cyclic permutations of $u,v,w$.

\subsubsection{Racah matrices and quantum $R$-matrices}

In fact, Vogel universality seems to survive quantization far beyond dimensions in the adjoint sector --
and thus Rosso-Jones formulas for torus knots.
Extension to arborescent calculus \cite{arbor}, for example, requires the knowledge of peculiar Racah (mixing) matrices
$U_{\rho\rho'}$, connecting
\be
\underbrace{(\mu \otimes \bar\mu)}_\rho\otimes \mu \longrightarrow \mu
\ \ \ \ \  {\rm and}  \ \ \ \ \
 \mu \otimes \underbrace{(\bar\mu\otimes \mu)}_{\rho'} \longrightarrow \mu
\ee
They indeed can be found in universal form -- so far this is done for a mixture of six representations in the square of $\mu$=Adjoint
(which is real, $\bar\mu=\mu$),
see \cite{MMracah}.

In addition to these mixing matrices,
arborescent calculus \cite{arbor} needs also {\it eigenvalues} of quantum $R$-matrices,
which are made from Casimir eigenvalues and can be expressed in the universal form  in the adjoint sector.

Vogel's universalization of generic $R$-matrices, including their Drinfeld's
{\it universal} expression (\ref{InvRT})
remains an open question.
It is even unclear to us, what can be special for the formula (\ref{InvRT}) in adjoint sector,
and to what extent it can be considered as made from the structure constants (\ref{JI}).
At the same time, there is a somewhat mysterious {\it eigenvalue conjecture} \cite{evconj},
which claims that $R$-matrices can be restored from their eigenvalues, and already played nicely in numerous applications.
Its interplay with Vogel universality is also an open question.

\subsection{Universal knot polynomials (Upols)}
\label{sec:upols}

In RT formalism HOMFLY and Kauffman polynomials are made from traces of products of quantum $R$ and $M$ matrices,
and are finally expressed through quantum dimensions, Casimirs and Racah matrices.
Since they all can be universalized, one can unify and lift these polynomials to {\it universal} ones \cite{MkrMM}.
We call them Upols.

The simplest set of Upols is provided by torus knots, where the  Rosso-Jones formula \cite{RJ,china,Che,AgSha,DMMSS}
is still applicable \cite{MkrMM}:
\be
U^{[m,n]}_R = \sum_{Q\in R^\otimes m} c_{_{RQ}} q^{\varkappa_R} {\cal D}_R
\label{URJ}
\ee
where $q^{\varkappa_R}$ are made from powers of $u,v,w$
and the new ingredient are just the Adams coefficients $c_{_{RQ}}$.
Fortunately they are just integer numbers, do not really depend on the algebra, and are described by a simple formula
in terms of symmetric-group characters, see the discussion at Section 2 of \cite{BiMi2}.
Examples for two and three strands $m=2,3$ and $R=Adj$ are given in \cite{MkrMM}, the case of $m=4$ is
described in \cite{BiMi2}.
In fact in the last case the answers are provided only for knots, similar formulas for links contain many more
representations.
Of course, calculation of universal dimensions of uirreps in $Adj^{\otimes m}$ is a separate scientific problem,
addressed in \cite{IP, uniformg4} and has its own levels of difficulty.
Remarkably, as we already mentioned in Section \ref{sec:dim}, the {\it controversial} representations from $Adj^{\otimes 4}$,
marked by the mathbbb font in \cite{uniformg4} drop out of Upols for {\it knots},
where even universal dimensions are still unknown.
The situation with links still remains obscure.
This issue is the main obstacle for claiming that (\ref{URJ}) is a totally clear formula for all $m$.

The vast variety of knots is handled by arborescent technique \cite{arbor} and its extension beyond arborescent
family in \cite{beyarbor}.
For the corresponding Upols one needs to know peculiar Racah matrices, in the simplest case of
$Adj^{\otimes 3}\longrightarrow Adj$ they are calculated in \cite{MMracah}.
The list of adjoint Upols for the knots from Rolfsen table, calculated in this way, is presented in
\cite{knotebook1}, namely at
\cite{knotebook2}.
The first white spot there (the first unknown adjoint Upol) is for $8_{18}$.
There are more white spots at $9$ and $10$ intersections.

Upols, colored by higher representations in adjoint sector were not systematically studied so far. Especially interesting can be Upols for mutants, like were the ordinary colored HOMFLY \cite{Morton,arbor}.

\subsection{Further deformations: hyper-, super- and Macdonald knot polynomials
\label{Mdknots}}

Double-graded knot invariants like HOMFLY \cite{HOMFLY} and Kauffman \cite{Kauffman} and their universal generalizations \cite{MkrMM}
possess at least a triple-grading generalization to Khovanov and Khovanov-Rozansky \cite{KhR} knot polynomials.
The lifting of the latter to non-(anti)symmetric representations is still a question,
still one can already ask a question about Vogel universality -- if, and to what extent it survives after this generalization.
So far no attempts were made
towards universalization of KR {\it complexes}, which remains a difficult technique
from the point of view of functional integrals and Chern-Simons theory.
The main approach is still more naive: one can try to relate Khovanov deformation to Macdonald one, introducing the notion of
{\it hyper}polynomials.
They are currently known \cite{Che,AgSha,DMMSS} for torus knots and links,
for the first attempts of generalization to knots, lying on other Riemann surfaces see \cite{ArtSh}.
The question of their relation to {\it super}polynomials \cite{GDR}, which are supposedly closer to KR,
is still open.
Therefore in practice the problem of universalization of KR polynomials is currently reduced to that of
Macdonald theory {\it per se}, with no true contact with knots --
the best hope at the moment would be existence of universal hyperpolynomials for torus knots.

Therefore what follows is a modest list of theses about universalization of Macdonald calculus.
This is a problem by itself, because Macdonald deformation leads us away from Dynkin's and Kac-Moody (affine)
Lie algebras and their ordinary quantum deformation(s), to the world of Yangians and DIM.
Since exact relation to hyper- and super- is not fully clear at this moment, we give the subject a separate name --
Macdonald knot polynomials.

\begin{itemize}
\item{} Dimensions of representations of Dynkin's algebras are made from Schur polynomials
-- and for representation in adjoint/$E_8$-sector they turn out to be universal.
One can define {\it Macdonald dimensions} in the same way from Macdonald polynomials \cite{DMMSS},
which themselves depend on the algebra (root system).
However, they {\it could} be universal only for ADE-algebras, but not for $B,C,G_2$ and $F_4$ \cite{BiMi}.
This is because beyond ADE the dual dimensions ${\cal M}_\lambda[q^{2\check{\rho}_k}]$
differ from the ordinary ones, ${\cal M}_\lambda[q^{2\rho_k}]$,
and even {\it before} Macdonald deformation the former are not universal.
At the same time, {\it after} this deformation only they remain factorized,
while universalization of non-factorized formulas is still a terra incognita.
In other words, Macdonald dimensions seem to break universality beyond ADE --
for other algebras factorized are the dual dimensions,
but they are not universal already before the $q,\mathfrak{t}$-transformation \cite{BiMi}.
In fact, the problem arises already at the level of Jack polynomials (also known as $\beta$-deformation,
where $\beta$ has nothing to do with Vogel's $\alpha,\beta,\gamma$.

\item{}
Within ADE subset Macdonald dimensions do have chances to be universal, for example (\ref{qdAdj}) is further deformed to \cite{BiMi}
\be
{\rm dim}^{\rm Mac}_{Adj} = \frac{\{\sqrt{u}vw\}\{\sqrt{v}uw\}\{\sqrt{w}uv\}}
{\{\sqrt{u}\}\{\sqrt{v}\}\{\sqrt{w}\}}
\cdot\frac{ \{uvw\}\{quvw\}  }{ \{\frac{q}{\mathfrak{t}}\,uvw\} \{\mathfrak{t}\,uvw\}  }
\label{MdAdj}
\ee
where now\footnote{
The two Macdonald parameters are  denoted by $q$ and $\mathfrak{t}$,
hopefully this $\mathfrak{t}$ will not be confused with Vogel's  $t=\alpha+\beta+\gamma$ in other parts of this paper.}
 $u=\mathfrak{t}^\alpha$, $v=\mathfrak{t}^\beta$, $w=\mathfrak{t}^\gamma$.
But in fact they are not, only the products ${\cal N}^\lambda_{\mu\nu}\cdot{\rm dim}^{\rm Mac}_\lambda$ can, see below.

\item{}
The present status of Macdonald deformation can be characterized by four statements:
\begin{itemize}
\item{} Universality, if at all preserved, remains only for ADE series.
\item{} Products of Macdonald polynomials do not quite reproduce product of representations,
when integer multiplicities (Littlewood-Richardson coefficients) $ n_\lambda^{\mu\nu}>1$ occur in
\be
\mu \otimes \nu = \oplus_\lambda n_\lambda^{\mu\nu} \cdot \lambda
\ee
the coefficients ${\cal N}_\lambda^{\mu\nu}$ in the expansion
\be
{\cal M}_\mu\cal M_\nu = \sum {\cal N}_\lambda^{\mu\nu} {\cal M}_\lambda
\ee
do not become sums of $ n_\lambda^{\mu\nu}$ factorized functions of $q$ and $\mathfrak{t}$.
Still ${\cal N}_\lambda^{\mu\nu}$ vanishes whenever $n_\lambda^{\mu\nu}=0$
\item{} It is not quite clear, how Macdonald {\it dimensions} \cite{DMMSS}
are represented in the universal form, like (\ref{MdAdj}) -- even for the ADE sector.
This problem looks simpler for "symmetric" dimensions like $D_{X_n}$ than for
asymmetric ones like $D_{Y_2(\alpha)}$.
Like in other cases, universal seem to be the quantities, which are seen in knot polynomials,
not in representation theory.
In this particular case un-separable-without-breakdown-of-universality are
Macdonald dimensions  ${\rm dim}^{\rm Mac}_\lambda:={\cal M}_\lambda[q^{2\rho_k}]$
and the Littlewood-Richardson
coefficients ${\cal N}^\lambda_{\mu\nu}$, which are also non-trivial functions of $q$ and $\mathfrak{t}$.
If anything is universal, it is a product ${\cal N}^\lambda_{\mu\nu}\cdot {\rm dim}^{\rm Mac}_\lambda$, not the two factors separately \cite{BiMM}.
This observation can also shed new light on the $\gamma$-factors in \cite{DMMSS}, even beyond adjoint representations.
See \cite{BiMi2} for current status of this statement.
\item{} In fact, all what we  know about dimensions is what they are for particular algebras — and all of them are defined to have $\alpha=-2$.
One usually restores the $\alpha$-dependence by requirement that formulas
are symmetric in $\alpha$,  $\beta$ and $\gamma$. However, the factorized quantities ${\cal N}^\lambda_{\mu\nu}\cdot{\rm dim}^{\rm Mac}_\lambda$ in the square of the adjoint do not obey this rule — a triple of representation is linked by the permutations of $u,v,w$, 
provided $u=q^{\alpha}$, $v=t^{\beta}$, $w=t^{\gamma}$,
i.e. these are {\it not} just the same as permutations of $\alpha$, $\beta$, $\gamma$.
In these new variables formula \eqref{MdAdj} also changes, to compensate for the changed definition of $u$.
However, it remains symmetric in $u,v,w$ and coincides with \eqref{MdAdj} at $\alpha=-2$. See [33] for detailed presentation of this story.
\end{itemize}
The first two items in this mini-list are old, while the last two are new.
One can consider them both pessimistically -- as a {\it problem} with Macdonald deformation
of universality -- or optimistically -- as a {\it sign} that it can be preserved, at least partly.
A possible way out can be provided by further generalization of Macdonald theory
on the lines of \cite{BatTs}, which preserves more of its properties, than generic Kerov
deformation \cite{Kerov} -- still can be broad/generous
enough to give more chances
to universality.

\item{} One can wonder about similar characteristics for Yangians and DIM,
which are  defined as non-Lie deformations of all simple Lie (super)algebras.
Appropriate quantities to look at are again somehow made from Jack and Macdonald polynomials,
and one can expect the same problems.
The problem is still  under investigation.

\item{} In fact there are additional problems for DIM case even with universality for just a single $A$-line:
while eigenfunctions of cut-and-join operators for $Y(\widehat{gl}_r)$ were provided by a nice family
of $r$-Uglov polynomials \cite{Uglov},
there seem to be problems with their further $t$-deformation to $T(\widehat{gl}_r)$
\cite{GMTprobs}.

\item{}
Significance of these deviations from universality is still unclear and can be easily ascribed to
insufficient understanding of Macdonald deformations beyond the ordinary ADE algebras.

\end{itemize}

\section{Conclusion
\label{conc}
}

This paper describes the scientific background behind the recent ITEP-JINR-YerPhI workshop
on Modern Trends in Vogel Theory, which was held in Dubna 21-23 April 2025.
It consisted of 6 lectures and discussion around them.

\bigskip

Aleksander Provorov presented his results with Alexey Isaev and Sergey Krivonos \cite{IP}
on the decomposition of powers of adjoint representation (up to 5)
and the search of universal formulas for emerging dimensions
with the help of the split-Casimir formalism.
One of the problems in this approach is that some representations
are not separated by the second split-Casimir, and some expressions
do not appear in the factorized form.
It is still unclear if this is a temporal technical problem of the formalism,
or it already reflects the ambiguities in universal formulas caused by divisors of zero --
what can make them well defined only modulo $P_{15}$ and $P_{22}$.

Ruben Mkrtchyan \cite{Mkr2} described the Diophantine equations, emerging from factorization requirement
for the quantum universal dimension of the adjoint representation,  eq.(\ref{qdAdj}).
This simply looking question appears related to Platon solids \cite{Khuda}
and leads to non-trivial pattern of points in the $\alpha,\beta,\gamma$ space,
not only located on the four Vogel's lines (i.e. zeros of $t\cdot P_{15}$)
but includes also two more lines and a number of discrete points.
The significance and the meaning of this observation is yet unclear.

Dmitry Khudoteplov gave a detailed lecture about Vogel's commutative $\Lambda$-algebra,
the (still open) problem of $\hat x_n$-induced basis in it and the zero-divisor(s).

Alexey Sleptsov then reviewed applications to knot theory and the problem of "extra"
Vassiliev invariants, non-extractable from HOMFLY, Kauffman and exceptional polynomials,
which are all trivialized on the $D_{17}$ diagram (related to the zero of $\hat t\cdot \hat P_{15}$).

Another lecture of R.Mkrtchyan was devoted to universalization of refined Chern-Simons theory
(i.e. its deformation from Schur to Macdonald sector),
which was supposed to go along the lines of \cite{MV, Mkrefined}, but faces some problems.

A parallel presentation of Ludmila Bishler \cite{BiMi} explained the origins of these problems,
at the level of Macdonald dimensions, which fail to factorize beyond the ADE algebras.

\bigskip

\bigskip

Our paper reviews all these current challenges, but puts them into a more general context
of the interplay between  Cartan-Dynkin-Kac and Vogel approaches to simple Lie algebras,
namely to classification of solutions of Jacobi identities.
Attention paid to Vogel theory and its close links to the fastly developing knot theory
gives hopes that these quadratic relations will be finally understood,
what would provide a reliable approach to many similar non-linear (classification) problems.
In fact, (\ref{JI}) is closely related to the no less celebrated bilinear Hirota equations,
which lie in the base of integrability theory and its profound applications in modern theory \cite{UFN3}.
Integrability is well known to be basically a part of Lie algebra theory --
and it is quite interesting to understand the role of universality in integrable systems.
At the same time, the hope is that Chern-Simons and knot theories will be also extended in the directions
of universality and beyond ordinary Lie algebras, including various deformations of
Jacobi identities -- not only to quantum groups but also to Yangians and DIM,
and, perhaps, much further.

In particular, there is a question, whether universality can be extended to Itzykson-Zuber integrals \cite{IZ,IZM,Hasib}
and further, to a full-fledged Yang-Mills theory,
not obligatory topological and also in even dimensions.
The funny restriction is that it would apply to adjoint sector only, where  there is no confinement,
at least in the usual sense of the area law.
Also in the Higgs theory adjoint Higgses manifest themselves very differently.
Perhaps, a physical meaning of universality is somehow related to this fact.
Interesting physics (confinement of quarks, monopole condensation and  true spontaneous symmetry breaking)
is non-universal and strongly depends on the gauge group, while the physically-not-so-relevant adjoint complement
is universal?
This is very different from the Chern-Simons situation, where the universal adjoint sector does not look so much different
from the fundamental one (Upols do not look too different from the ordinary HOMFLY).
Perhaps, we overlook something,
and there is much more to {\it understand}, not just calculate about knots?
Alternatively, confinement of adjoint gluons can happen to be not so different from confinement of quarks,
as it is often assumed?

\

It would be interesting to see, which of these hopes will get realized, and how they will do so.

\section*{Acknowledgements}

We are indebted to the organizers and participants of the workshop,
especially to
L.Bishler, A.Isaev, D.Khudoteplov, S.Krivonos, E.Lanina, A.Mironov, S.Mironov,
R.Mkrtchyan, An.Morozov, A.Popolitov, A.Provorov and N.Tselousov.

Our work is partly funded within the state assignment of IITP RAS. It is partly supported by the grants of the Foundation for the Advancement of Theoretical Physics and Mathematics ``BASIS".

\end{document}